\documentclass[]{aastex631}

\begin{document}

\title{THE SUBPARSEC-SCALE STRUCTURE AND EVOLUTION OF CENTAURUS A. III. 
A Multi-Epoch Spectral And Polarimetric VLBA Study}

\author[0000-0003-3165-6785]{S. Prabu}
\affiliation{International Centre for Radio Astronomy Research, Curtin University, Bentley, WA 6102, Australia}
\author[0000-0002-8195-7562]{S.J. Tingay}
\affiliation{International Centre for Radio Astronomy Research, Curtin University, Bentley, WA 6102, Australia}
\author[0000-0003-2506-6041]{A. Bahramian}
\affiliation{International Centre for Radio Astronomy Research, Curtin University, Bentley, WA 6102, Australia}
\author[0000-0003-3124-2814]{J.C.A. Miller-Jones}
\affiliation{International Centre for Radio Astronomy Research, Curtin University, Bentley, WA 6102, Australia}
\author[0000-0002-2758-0864]{C.M. Wood}
\affiliation{International Centre for Radio Astronomy Research, Curtin University, Bentley, WA 6102, Australia}
\author[0000-0002-3968-3051]{S.P. O'Sullivan}
\affiliation{Departamento de Física de la Tierra y Astrofísica \& IPARCOS-UCM, Universidad Complutense de Madrid, 28040 Madrid, Spain}



\begin{abstract}
The Centaurus A radio galaxy, due to its proximity, presents itself as one of the few systems that allow the study of relativistic jet outflows at sub-parsec distances from the central supermassive black holes, with high signal to noise. We present the results from the first multi-epoch spectropolarimetric observations of Centaurus A at milliarcsecond resolution, with a continuous frequency coverage of $4.59-7.78$\,GHz. Using a Bayesian framework, we perform a comprehensive study of the jet kinematics, and discuss aspects of the jet geometry including the jet inclination angle, jet opening angle, and the jet expansion profile. We calculate an upper limit on the jet's inclination to the line of sight to be $<25^{\circ}$, implying the lower limit on the intrinsic jet speed to be $0.2$\,c. On the observed VLBA scales we detect new jet components launched by the central engine since our previous study. Using the observed frequency-dependent core shift in Centaurus A, we find the jet to have reached constant bulk speed and conical outflow at the regions probed by the base of the jet at $7.78- 4.59$\,GHz, and we also estimate the location of the central black hole further upstream. Through polarimetric analysis (by applying RM synthesis for the first time on VLBI data), we find evidence to suggest the possible onset of acceleration towards the leading edge of Centaurus A's subparsec-scale jet studied here. 
\end{abstract}



\section{Introduction} \label{sec:intro}
The gravitational potential energy of material falling onto a supermassive black hole (SMBH), some of which is released as outflowing jets is one of the most powerful physical process in the Universe \citep{hughes1991beams}. Some fraction of the infalling material is collimated into jet outflows very close to the black hole, thought to be due to the action of toroidal magnetic fields of the accretion disk \citep{blandford1982hydromagnetic} and/or due to the action of a rotating black hole on its environment \citep{blandford1977electromagnetic}, giving rise to a spine and sheath jet structure. The collimated outflow is expected to be accelerated at sub-parsec to parsec distances from the black hole \citep{vlahakis2004magnetic} before again being slowed down at 10-100 kiloparsec scales due to interaction with the intergalactic medium. In this paper, we present the results from the third part of a series of Very Long Baseline Array (VLBA) studies of the sub-parsec structure and evolution of the jet in the Centaurus A (PKS 1322-427, NGC 5128) radio galaxy. NGC 5128 hosts a supermassive black hole of mass  $(5.5 \pm 3) \times 10^{7} M_{\odot}$ \citep{neumayer2010supermassive} that powers the radio jet outflow studied here. Due to its proximity  \citep[$3.4$\, Mpc,][]{israel1998centaurus}, Centaurus A is one of the few ideal targets to study the launching and collimation of jet outflows at sub-parsec scales. At its distance, an angular size of $1$\,mas translates to a linear size of approx. $0.018$\,pc $\approx 1.7 \times 10^{3}$ gravitational radii ($R_{g}$ = $GM/c^{2}$).\\

The compact radio nucleus of Centaurus A hinted at by the observations by \cite{1974ApJ...194L.135K}, \cite{1971ApJ...170L..11W}, and \cite{1973NPhS..245...83P}, was first revealed with very long baseline interferometry (VLBI) visibility measurements by \cite{1983ApJ...266L..93P}, using three antennas in Australia and South Africa at 2.3 GHz.  The first VLBI observations capable of imaging the nucleus came soon after, with \cite{1989AJ.....98...27M} producing the first model of the sub-parsec scale Centaurus A jet and nucleus at 2.3 and 8.4 GHz. This led to Centaurus A becoming a regular target for Southern Hemisphere VLBI observations in the 1990s, primarily at 8.4 GHz, which provided sufficient angular resolution to resolve the sub-parsec-scale jet and took advantage of the strongly inverted spectrum of the compact nucleus \citep{1974ApJ...194L.135K}.  Via these observations, Centaurus A was shown to be one of the few radio galaxies at the time to display a convincing parsec-scale receding-jet \citep{jones1996discovery}.\\

\begin{figure*}[h]
\begin{center}
\includegraphics[width=\linewidth]{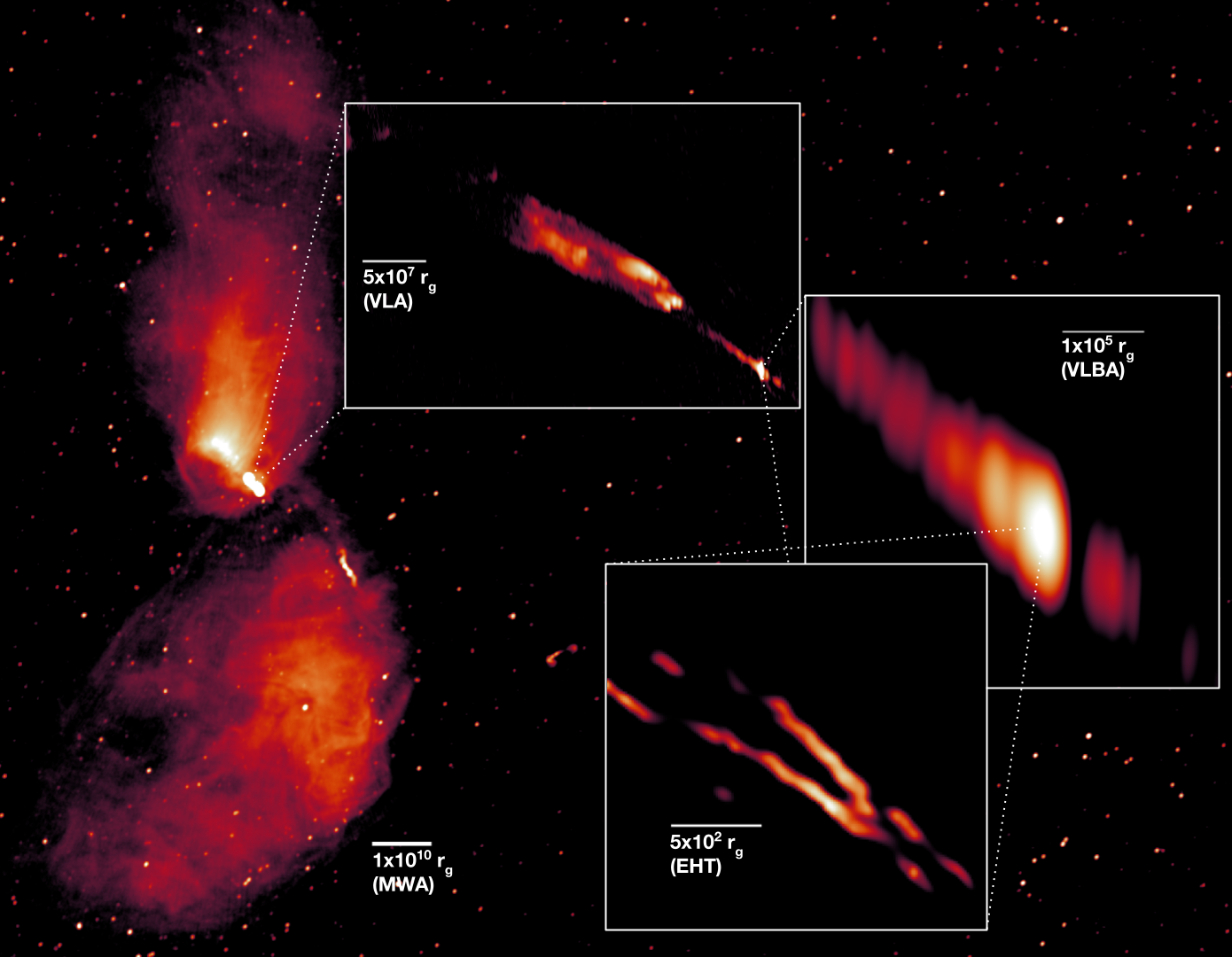}
\caption{A series of zoom-in plots shows the jet outflow from Centaurus A's central engine at various spatial scales. On on the largest scales, we show the 10-100 kiloparsec jet \citep{2022NatAs...6..109M} imaged with the Murchison Widefield Array (MWA). The Very Large Array (VLA) image of Centaurus A shows the kilo-parsec scale jet \citep{2003ApJ...593..169H}, and the Event Horizon Telescope image shows the jet outflow very close to the black hole \citep{janssen2021event}. The VLBA jet studied in this work is the intermediate region between the EHT image and the VLA image. The $R_{g}$ shown is along the plane of the sky (not corrected for inclination angle), and are calculated using a mass of $5.5 \times 10^{7} M_{\odot}$ \citep{neumayer2010supermassive} and distance of $3.4$\, Mpc \citep{israel1998centaurus}.}
\label{fig:zoomin}
\end{center}
\end{figure*}

Using eight years of Southern Hemisphere VLBI observations from 1988-1995, \cite[][ hereafter {\tt PAPER I}]{tingay1998subparsec} undertook the first systematic multi-epoch examination of the structure and evolution of the sub-parsec-scale jet, detecting multiple components (labeled as C1, C2, and C3) along the approaching jet, appearing to move away from the nucleus. The jet components displayed a wide range of apparent speeds spanning from the negligible motion of C3 during the 8 years of observation, to the C1 component moving at an apparent speed of $0.1$\,c away from the nucleus. The jet was observed at a position angle of $48^{\circ}$ (measured East of North), with the North-Eastern jet being Doppler boosted towards us. Using the brightness ratio of the approaching jet and the receding jet \citep{jones1996discovery}, {\tt PAPER I} determined the jet's inclination with respect to the line of sight to be between $50^{\circ} - 80^{\circ}$. \\

In the early 1990s, the Very Long Baseline Array (VLBA) came into existence, providing the first dedicated and built-for-purpose VLBI array.  Even at the $-45^{\circ}$ declination of Centaurus A, the VLBA obtains good (u,v) coverage for imaging studies.  Over the next decade, the VLBA became the workhorse for Centaurus A VLBI observations, exploring different aspects of the jet outflow. \cite{tingay2001subparsec} (hereafter {\tt PAPER II}) continued observing Centaurus A using the VLBA during the period 1995-2000 over multiple frequencies, and confirmed the slow-moving C3 component to be a quasi-stationary extension of the nucleus. {\tt PAPER II} also detected very slow apparent motion of components in the receding jet, as expected for a jet inclined at $50^{\circ} - 80^{\circ}$. While \cite{1998ASPC..144..131V} and \cite{2005Ap&SS.295..249V} used the VLBA to study molecular absorption towards the sub-parsec scale jet and nucleus of Centaurus A, \cite{tingay2001estimates} used 2.3, 4.8, and 8.4 GHz VLBA observations to examine the spectral distribution of the jet outflow. Towards the higher frequencies, \cite{1997ApJ...475L..93K} used the VLBA at 43 GHz to obtain the highest angular resolution possible for Centaurus A at that time, placing an upper limit on the size of the compact nucleus to be $\le 0.01$ pc. The polarisation properties of the inverted spectrum nucleus were examined using the VLBA by \cite{2005ASPC..340..189M} at 15 GHz, showing no evidence for polarised emission. \\

Simultaneously, other very high angular resolution VLBI observations were also conducted using the first space VLBI mission, VLBI Space Observatory Programme \citep[VSOP;][]{fujisawa2000large, horiuchi2006ten}, as well as by the Tracking Active galactic Nuclei with Austral Miliarcsecond Interferometery program \citep[TANAMI;][]{muller2011dual, 2013arXiv1301.4384M, muller2014tanami}.  The TANAMI observations first revealed structures in the jet at a fixed distance from the nucleus that had previously been hinted at as sites of jet disturbance \citep[][and {\tt PAPER I}]{1989AJ.....98...27M}.  The structures had a ``tuning fork'' morphology and were suggested to be the result of the interaction of the jet with a star.  Interestingly, jet-star/cloud interactions have been suggested to explain jet structures in Centaurus A on the kiloparsec-scale by \cite{2003ApJ...593..169H} and \cite{2009AJ....138..808T}.  The proximity of Centaurus A may make it the only radio galaxy for which such physical scenarios can be probed and tested with detailed observations. \\

Even higher angular resolution VLBI imaging has been performed recently, using the Event Horizon Telescope (EHT), exploiting 228 GHz observations to examine the jet launching and collimation processes 200 $R_{g}$ from the supermassive black hole \citep{janssen2021event}. Today, Centaurus A remains to be one of the few radio galaxies where the outflow of relativistic jets has been explored over a large range of spatial scales at radio frequencies \citep{janssen2021event, tingay1998subparsec, tingay2001subparsec, muller2011dual,2009ApJ...707..114F, anderson2018extraordinary, 2022NatAs...6..109M} (see Figure \ref{fig:zoomin} for radio emission at different scales) and also X-rays \citep{2002ApJ...569...54K, 2022ApJ...932..104R, 2011A&A...531A..70B, ehlert2022limits}. To contrast with the high-resolution VLBI observations discussed above, the Murchison Widefield Array \citep[MWA;][]{tingay2013murchison}, with large-n array configuration and baseline lengths $<6$\,km has revealed the large scale kinematic feedback of the jet outflow in Centaurus A \citep[][see Figure \ref{fig:zoomin}]{2022NatAs...6..109M}. \\

In this paper, {\tt PAPER III} of our series, we report continued observations of Centaurus A across 12 epochs using the VLBA, from early to late 2013 at a cadence of weeks which, for the first time, systematically probe short timescales, motivated by the rapid variability occasionally observed at the ``tuning fork''. The observations had continuous frequency coverage from $4.59 - 7.78$\, GHz, allowing a much more detailed examination of the spectral behaviour of the jet and nucleus than was possible from previous observations \citep{tingay2001estimates}. The observations were configured to search for any polarised emission at sub-parsec distances from Centaurus A's SMBH, at frequencies that probe the optically thin emission of the jet, which would have been missed in prior searches for polarised emission at much higher frequencies \citep{2005ASPC..340..189M}.\\

This paper is structured as follows. In Section \ref{sec:datareduction} we briefly explain the observation setup and the data processing procedures. In Section \ref{sec:results}, we present the results from our jet kinematics analysis, spectral analysis, and polarisation analysis. We discuss our results and conclude in Sections \ref{sec:discussion} and \ref{sec:conclusion}, respectively. \\

\section{OBSERVATIONS AND DATA REDUCTION} \label{sec:datareduction}
Centaurus A was observed using the VLBA for 12 epochs between 2013-01-27 and 2013-08-23 under project ID BO043. The observation campaign made use of the wide-bandwidth capability of the VLBA to obtain continuous frequency coverage from $4.59 - 7.78$\, GHz, split into 12 bands of $256$\, MHz bandwidth each. All four polarisations were recorded (RR, LL, RL, and LR), and no phase-reference sources were used, as Centaurus A is bright enough to be self-calibrated. Of the twelve epochs, we only present here the results from eight epochs (see Table \ref{tab:observations}), as the remainder had very poor uv-coverage (either due to less than five antennas being available or due to extensive flagging that had to be performed).\\

\begin{table*}[h!]
    \caption{List of observations.}
    \centering
    \begin{tabular}{@{}lcccc@{}}
    Obs ID & Start (UTC)   &  Antennas available (before flagging) & Ref ant & Comment \\
           &  Stop (UTC)      &  Antennas available (after flagging) &  & (Reason) \\
    \hline \hline
    BO043A & 01/27/2013 09:15:00   & BR, FD, HN, KP, LA, MK, NL, OV, PT, SC  & KP & used here\\
    & 01/27/2013 15:12:00  & FD, KP, LA, NL, OV, PT   & & \\
    BO043B & 02/11/2013 08:16:00   &  BR, FD, HN, KP, LA, MK, NL, OV, PT, SC & KP & used here\\
    & 02/11/2013 14:13:00  & FD, KP, LA, NL, PT   & & \\
    BO043C & 03/06/2013 06:45:00   & BR, FD, HN, KP, LA, NL, OV, PT, SC  & KP & used here\\
    & 03/06/2013 12:45:00  & FD, KP, LA, NL, OV, PT   & & \\
    BO043D & 04/06/2013 04:43:00   & FD, HN, KP, LA, MK, NL, OV, PT, SC    & KP & used here\\
    & 04/06/2013 10:43:00  & FD, KP, LA, NL, OV, PT  & & \\
    BO043E & 04/23/2013 03:37:00   &  BR, FD, HN, KP, LA, MK, NL, OV, PT, SC & KP & used here\\
    & 04/23/2013 09:36:00  & FD, KP, LA, NL, OV, PT  & & \\
    BO043F & 05/11/2013 02:26:00   & BR, FD, HN, KP, LA, MK, NL, OV, PT, SC   & KP & used here\\
    & 05/11/2013 08:26:00  &  FD, KP, LA, NL, OV, PT & &   \\
    BO043G & 06/11/2013 00:24:00  & BR, HN, KP, LA, MK, NL, OV, SC &  & not used\\
    & 06/11/2013 06:24:00   & KP, LA, NL, OV & & (poor uv-coverage)\\
    BO043H & 06/25/2013 23:25:00   & HN, KP, LA, MK, NL, OV, PT, SC &  & not used\\
    & 06/26/2013 05:25:00  & KP, LA, NL, OV, PT & & (7/8 IFs down) \\
    BO043I & 07/06/2013 22:42:00   & BR, HN, KP, LA, MK, NL, OV, PT, SC & KP  & used here$^{*}$\\
    & 07/07/2013 04:42:00  & KP, LA, NL, OV, PT  & &  \\
    BO043J & 07/29/2013 21:11:00   & BR, HN, KP, LA, MK, NL, OV, PT, SC  &  & not used\\
    & 07/30/2013 03:09:00  & KP, LA, NL, PT & & (poor uv-coverage)\\
    BO043K & 08/09/2013 20:28:00   & BR, HN, KP, LA, MK, NL, OV, PT, SC &  & not used\\
    & 08/10/2013 02:26:00  & LA, NL & & (poor uv-coverage)\\
    BO043L & 08/23/2013 19:33:00   & BR, FD, HN, KP, LA, MK, NL, OV, PT, SC & FD & used here\\      
    & 08/24/2013 01:30:00  & FD, KP, LA, NL, OV, PT & & \\
    \hline \hline
    \end{tabular}
    \label{tab:observations}
     {\raggedright $^{*}$Epoch BO043I had very large holes in UV coverage, often resulting in negative features in the images. With careful flagging, we are able to use the $7.78$\, GHz observation to study jet kinematics (despite having no detection of the C1 component and a very low signal-to-noise detection of the C2 component). However, due to applying further strict limits on the UV range for the spectral analysis and RM-synthesis analysis (in order to ensure all frequencies probe similar spatial scales), this epoch was not usable for either of the analyses. \par}
\end{table*}

\begin{figure*}[h]
\begin{center}
\includegraphics[width=\linewidth]{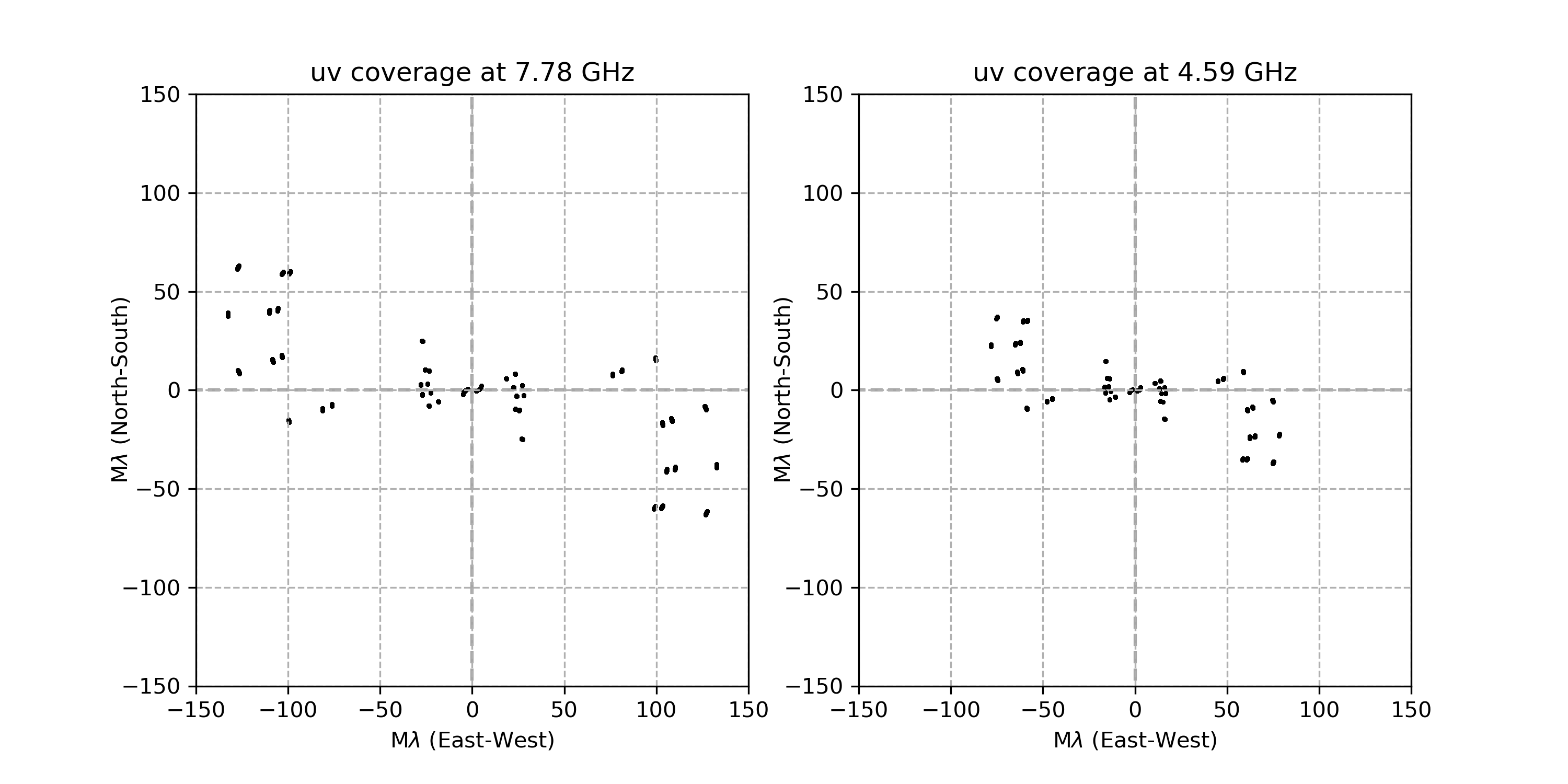}
\caption{The uv-coverage at the highest and the lowest frequency bands for the epoch BO043A, after having flagged BR, HN, MK, and SC VLBA stations. The baseline lengths are measured in mega-wavelengths along the East-West and North-South directions.}
\label{fig:uvcoverage}
\end{center}
\end{figure*}

The observations were processed using the Astronomical Image Processing System \citep[{\tt AIPS};] [version 31 DEC22]{Greisen2003} following the standard procedure for VLBI polarimetry experiments, as described in the {\tt AIPS} cookbook\footnote{\url{http://www.AIPS.nrao.edu/cook.html}} and EVN MEMO $\# 78$\footnote{\url{https://www.brandeis.edu/departments/physics/astro/pdfs/evnmemo78.pdf}}. Due to the low declination of Centaurus A, {\tt PAPER I} (and {\tt PAPER II}) flagged the Brewster and Hancock antennas in order to avoid residual ionospheric delays from low-elevation scans. The previous work also found flagging Mauna Kea and Saint Croix antennas was needed to produce reasonable uv-coverage to minimise imaging artifacts due to large holes in the uv-plane. Hence, in order to keep the observations in this paper comparable to the previous work, we follow the same antenna flagging strategy as in {\tt PAPER I/II}. Having performed the above-mentioned flagging, we show the resultant uv-coverage for the highest and the lowest frequency bands for epoch BO043A in Figure \ref{fig:uvcoverage} \\

\begin{figure*}[h]
\begin{center}
\includegraphics[width=\linewidth]{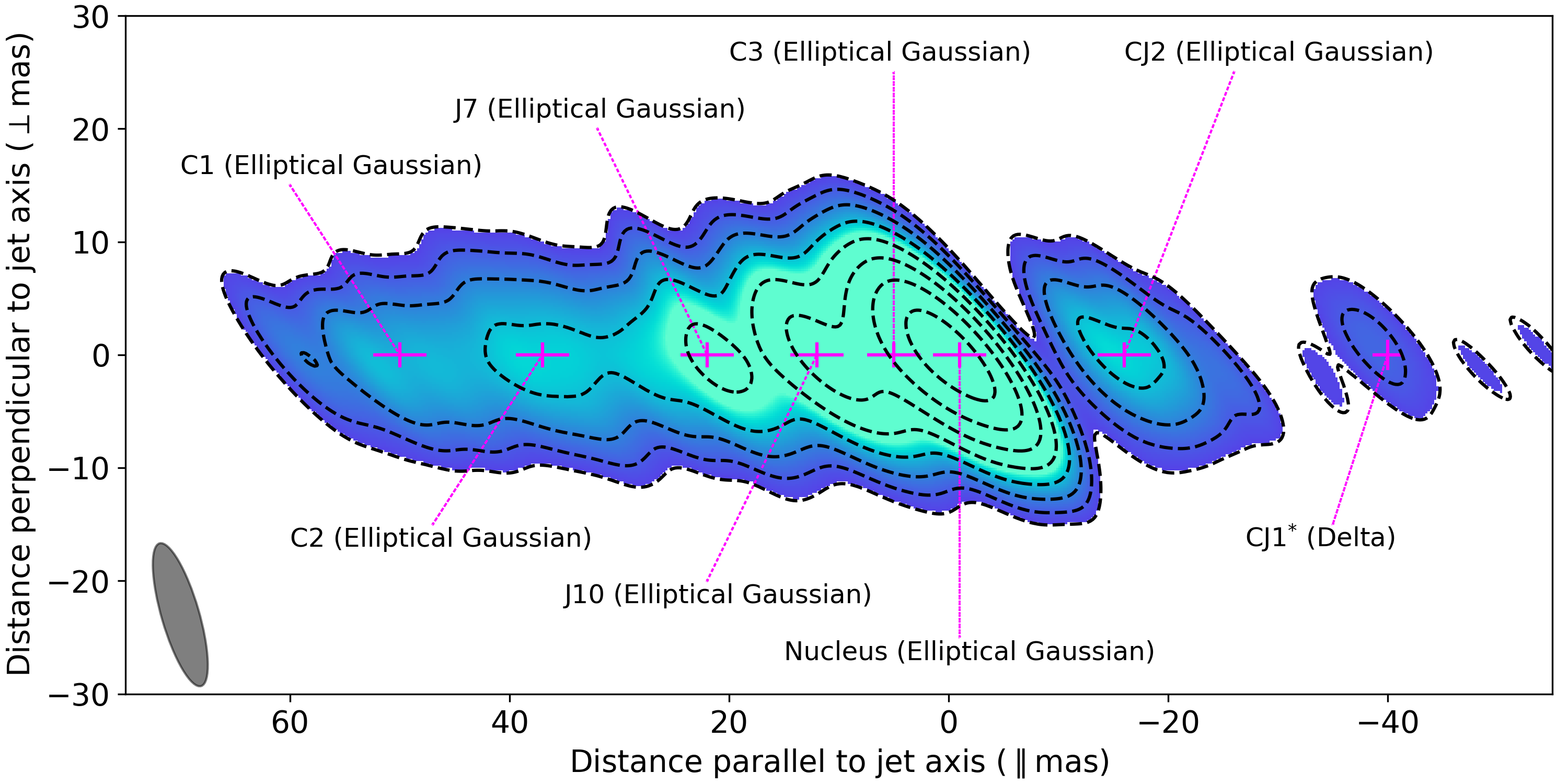}
\caption{A stacked $7.78$\,GHz naturally weighted image of Centaurus A during the eight epochs listed in Table \ref{tab:observations}. The distances are measured from the base of the jet at $7.78$\,GHz. We also annotate the different components detected in the approaching and receding jets along with the choice of model component used to represent the feature. The previously detected VLBA components are named using the same names used in {\tt PAPER I/II} (as C1, C2, C3, and CJ2), and the more recently launched jet components are named using the names used in TANAMI (J7 and J10). The contour levels increase as, $noise \times \sqrt{2}^{n}$, where $n=3, 5, 7, 9, ...$. The restoring beam size is shown in grey at the bottom left, and the rms noise in the image is $2.7$ mJy/beam. All pixels below $3\times noise$ have been masked to be white for better illustration.}
\label{fig:diagram}
\end{center}
\end{figure*}

We started by removing ionospheric delays, delays from incorrect Earth Orientation Parameters (EOPs), and sampler bias using the {\tt AIPS} procedures {\tt VLBATECR}, {\tt VLBAEOPS}, and {\tt VLBACCOR}, respectively. Using the stored antenna temperatures, we then performed a-priori amplitude calibration using {\tt VLBAAMP}. Our observations spanned a few hours, scanning the target and different calibrators over a wide range of parallactic angles. Hence, {\tt VLBAPANG} was run on the uv dataset to remove the effects of changing parallactic angle on the measured cross-hand polarisation (RL and LR). We then used pulse calibration scans of J1316-3338 to align the phases across all intermediate frequencies (IFs) using {\tt VLBAPCOR}, followed by application of bandpass calibration using {\tt VLBABPSS}. The fringes were then aligned across all scans by performing a global fringe-fit (self-calibration) using the {\tt FRINGE} task, and this step brought the brightest feature in Centaurus A to the phase-center of the observation (and hence the absolute phase/position of Centaurus A is lost). As we also intended to perform polarisation measurements, we then removed the delays in the cross-hand polarisations using {\tt RLDLY}. All of these steps were done in a semi-autonomous manner using {\tt AIPS RUN} files that can be obtained from our attached supplementary material. Having applied the above calibrations, the calibrated uv-data of J1407+2827 was exported to {\tt difmap} \citep{shepherd1997difmap} in order to develop a source model, which is later used to determine the instrumental leakage (D-terms). \\

\begin{table*}[h!]
    \caption{Proper Motion of fit components. The medians and 1-sigma credible intervals of the fit parameters are tabulated below. }
    \centering
    \begin{tabular}{@{}lcc@{}}
    Comp name & Apparent speed (mas/year) [$\beta_{app}$] & $T_{ejection}$ (years)\\
    \hline
    C1  & 1.65 $\pm$ 0.19 [0.11c $\pm$ 0.01c] & 1981.45 $\pm$ 2.44\\
    C2 & 1.55 $\pm$ 0.23 [0.10c $ \pm$ 0.02c] & 1988.68 $\pm$  1.06\\
    J7  &1.84 $\pm$ 0.14 [ 0.12c $\pm$  0.01c] & 2002.07 $\pm$ 0.57\\
    J10  &2.81 $\pm$ 0.06 [0.19c $\pm$  0.01c] &  2008.69 $\pm$  0.11\\
    C3  & -0.02 $\pm$ 0.14 [-0.00c $\pm$ 0.01c] &\\
    CJ2 &  0.19 $\pm$ 0.15 [0.01c $\pm$ 0.01c] &\\
    \hline
    
    \end{tabular}
    \label{tab:kinematics}
\end{table*}

In {\tt difmap}, we self-calibrated and built a model of J1407+2827 using Gaussian  and delta function components. The {\tt difmap} model file of J1407+2827 was then loaded back into {\tt AIPS}, and was used to self-calibrate the original uv-data of J1407+2827 using {\tt CALIB}\footnote{Note that {\tt difmap} assumes $I=LL/2=RR/2$ during self-calibration and can produce spurious leakage across polarisations. Hence, we use {\tt CALIB} to self-calibrate the data prior to running {\tt LPCAL}, as it accounts for any potential Stokes Q emission appropriately.}. As J1407+2827 is an unpolarised source, any cross-hand emission in scans of J1407+2827 would be due to instrumental leakage. {\tt LPCAL} was run on the self-calibrated dataset of J1407+2827, and the AN table produced containing the D-terms was exported to the main multi-source uv dataset (using {\tt TBOUT} and {\tt TBIN}). Using {\tt SPLIT} (with the dopol=3 argument used in order to apply the D-terms), we separated the calibrated uv dataset for Centaurus A and J2202+4216, the Electic Vector Position Angle (EVPA) calibrator, from the multi-source dataset. The Centaurus A data split was then further processed differently based on the science case (e.g., jet kinematics, spectral index maps, etc) and is further discussed in the corresponding sections below. In order to set the absolute phase of the cross-hands, we processed the VLA observation of J2202+4216 that was closest to our observation campaign (2014-01-12, as no VLA observation was performed in 2013), following the steps provided in the CASA VLA tutorial\footnote{\url{https://casaguides.nrao.edu/index.php/CASA_Guides:Polarization_Calibration_based_on_CASA_pipeline_standard_reduction:_The_radio_galaxy_3C75-CASA5.6.2}}. We obtained a polarisation angle of $-62.8^{\circ}$ at $5.0$\,GHz and $-2.19^{\circ}$ at $8.4$\,GHz for J2202+4216 on 2014-01-12. Using a pre-2012 calibration solution published by NRAO\footnote{\url{http://www.vla.nrao.edu/astro/evlapolcal/index.html}} for J2202+4216 and our obtained values, we performed 2D interpolation (in time and wavelength squared) to obtain the calibrator's polarisation angle for individual epochs and bands. For all our polarisation analysis, we performed image-based EVPA corrections to the observed complex quantities. We define the EVPA correction as $\chi_{corr} = \chi_{true} - \chi_{obs}$, where $\chi_{true}$ is J2202+4216's EVPA as measured by the VLA and $\chi_{obs}$ is J2202+4216's EVPA measured in VLBA observations prior to EVPA correction. We then applied the corrections to every IF and epoch of Centaurus A by multiplying the observed polarised intensity ($Q + iU$) by $e^{2i\chi_{corr}}$. \\

\section{RESULTS} \label{sec:results}

\subsection{JET KINEMATICS} \label{sec:results:jetKinematics}
We study the sub-parsec structure and evolution of Centaurus A using the highest frequency band ($7.78$\,GHz) available within our VLBA campaign. The choice for using the highest frequency band for the study of jet kinematics is two-fold; first, it provides the highest resolution images of the Centaurus A jet, enabling us to perform precise astrometry of jet components. Secondly, the use of the highest frequency band also ensures continuity between the jet kinematics studied in this paper with our previous work ({\tt PAPER I/II}). {\tt difmap} is used to self-calibrate and model fit the Centaurus A uv dataset obtained from {\tt AIPS}. The compact nucleus and the jet components are modeled as elliptical Gaussians in {\tt difmap}. Occasionally, the jet components along the leading edge of the approaching and receding jets were detected with reduced signal-to-noise and hence were modeled using delta components instead of an elliptical Gaussian.  In order to keep our astrometry error estimation consistent with the approach taken by {\tt PAPER I/II}, we use the major/minor axis of the fit elliptical Gaussian projected along the known jet axis to be a conservative estimate of the error associated with the component's astrometry. For delta components, we use the full-width at half maximum of the synthesized beam as the astrometry error. \\

Upon inspecting the images made at different frequencies and epochs, we consistently find five distinct components along the approaching jet, and one or more components in the receding jet direction, as shown in Figure  \ref{fig:diagram}. Having accounted for the proper motion of the components detected by {\tt PAPER I/II}, we continue to detect all the previously detected components along the approaching jet (C1, C2, and C3), along with two new components. The observations used in this work ({\tt PAPER III}), uses VLBA observations from 2013, more than a decade after the observations used in {\tt PAPER I/II}. However, the TANAMI study of Centaurus A \citep{muller2011dual, muller2014tanami} uses observations (2007-2011) that are closer in time to the 2013 VLBA observations used here and hence captures the ejection of more recent jet components. We find our two new approaching jet components to spatially coincide with the J7 and J10\footnote{By comparing the spatial location of this VLBA component with the TANAMI campaign, it is likely to be either the J9 or the J10 component. The J9 component consistently has flux density measurements of $<0.27$\,mJy in the TANAMI campaign, and hence, of the two, it is more likely to be the J10 due to it having similar flux densities to what we measure in our VLBA images. However, as J9 and J10 are close in proximity and have similar velocities, it is also possible we are seeing a low-resolution amalgamation of the two components in our VLBA images. } components detected by TANAMI with similar flux densities. Note that for epoch BO043I, we do not detect the component C1, and C2 was detected with reduced signal to noise\footnote{we attribute this to only five unflagged antennas being available.} and hence was modeled as a delta function instead of an elliptical Gaussian.\\

In the receding jet direction, we continue to detect the CJ2 component from {\tt PAPER II}, along with one additional component detected further downstream, which we call CJ1$^{*}$ in the remainder of the paper (since it may or may not be associated with the CJ1 component detected in {\tt PAPER II}, and is further discussed below). Unlike the bright CJ2 component, CJ1$^{*}$ is only detected in some of the epochs, and with very low signal-to-noise (and hence is modeled as a simple delta function). In many of the epochs, the CJ1$^{*}$ component was detected at different distances ($30 - 50$\,mas) from the nucleus, hence it is very likely that we are detecting different faint receding jet components at different epochs (due to changing uv coverage between observations). Hence, due to the large variations in our CJ1$^{*}$'s position, we refrain from associating it with the previously detected CJ1 component in {\tt PAPER II}. \\

\begin{figure*}[h]
\begin{center}
\includegraphics[width=\linewidth]{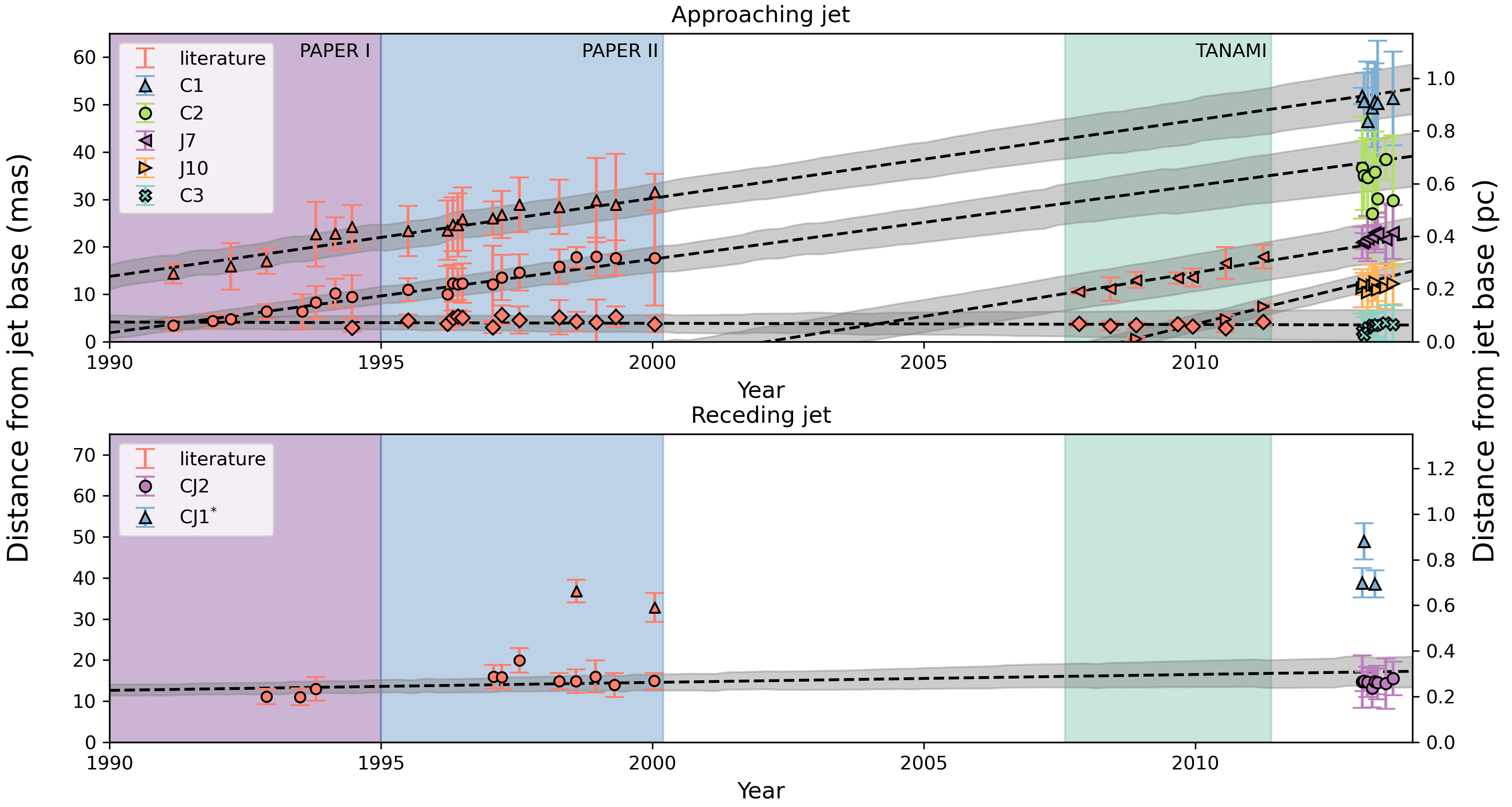}
\caption{Astrometry and proper motion of the jet components detected along the approaching jet (top panel) and the receding jet (bottom panel). The distances of the jet components are measured from the base of the jet at $7.8$\,GHz (nucleus). In orange, we show data for the components obtained from the literature ({\tt PAPER I}, {\tt PAPER II}, and TANAMI), along with their errors. The black dashed line is the best fit proper motion obtained from using our model fit components along with the ones already published in the literature. The grey shaded area is the $68\%$ high-density interval of our proper motion fit. }
\label{fig:properMotion}
\end{center}
\end{figure*}

\begin{figure*}[h]
\begin{center}
\includegraphics[width=\linewidth]{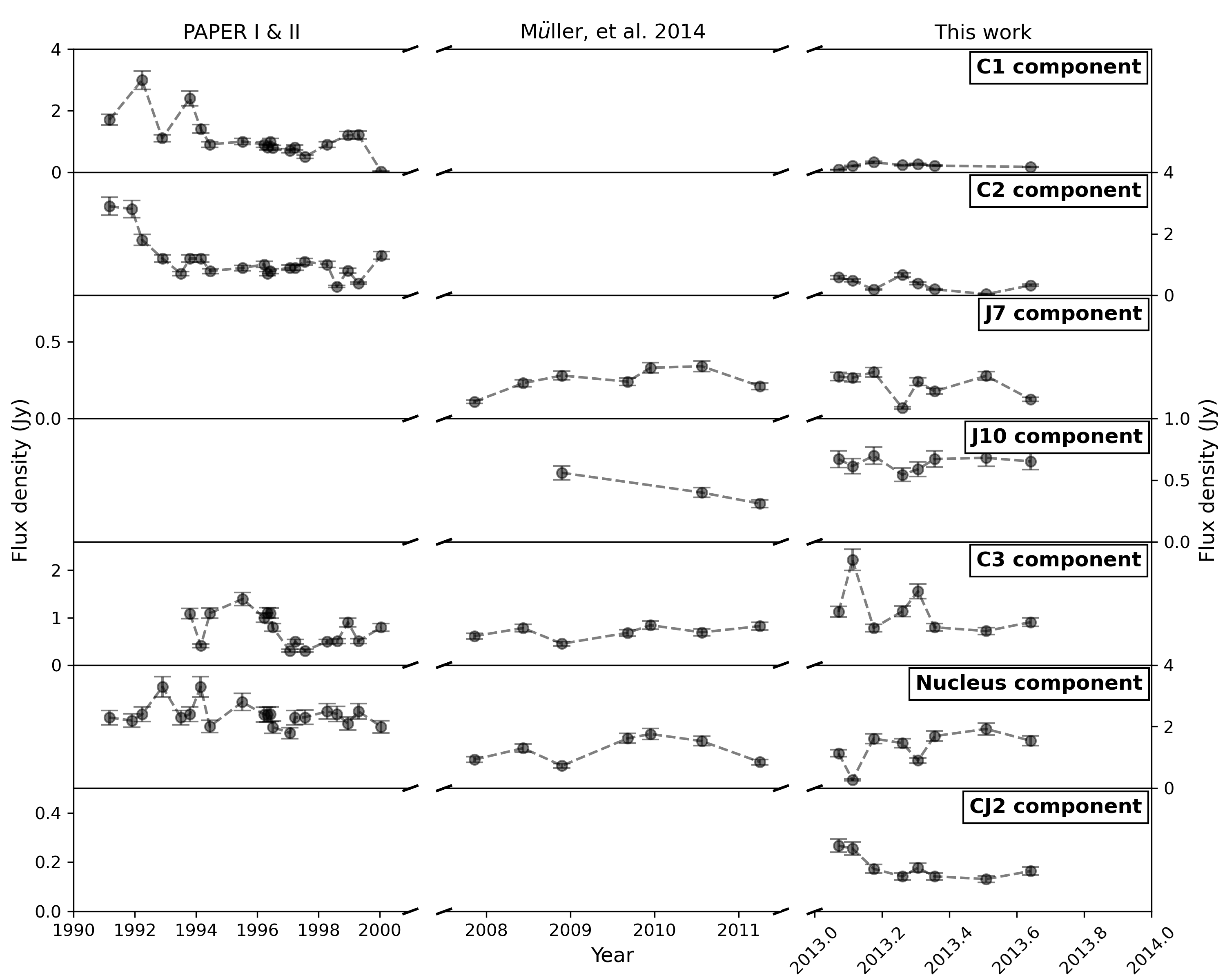}
\caption{The light curves for all the model components. We also show the historical flux density evolution for the identified components from previous studies. }
\label{fig:lightCurve}
\end{center}
\end{figure*}

Having associated the components detected in the literature with those detected in this work, we proceed to analyse the kinematics of these components to estimate their apparent speeds and ejection dates. Since neither {\tt PAPER I} nor {\tt PAPER II}  detected any significant acceleration of components within the subparsec distances from Centaurus A's black hole, we use a Bayesian framework (described further in the following paragraph) that fits ballistic trajectories to these components. In the top and bottom panels of Figure \ref{fig:properMotion}, we show our measured distances of the jet components from the compact nucleus, as it represents the base of the jet (i.e, the optical depth $\tau \sim 1$ surface) at $7.8$\, GHz. As we measure the distances of these components from the nucleus, the nucleus component's astrometry error is added in quadrature to the individual distance measurements. The light curve of the model components is shown in Figure \ref{fig:lightCurve}, along with the values previously published in the literature. We use $10\%$ of the flux density added in quadrature to the image noise as the error estimates for our light curve, as often done in the VLBI literature. \\

\begin{figure*}[h]
\begin{center}
\includegraphics[width=\linewidth]{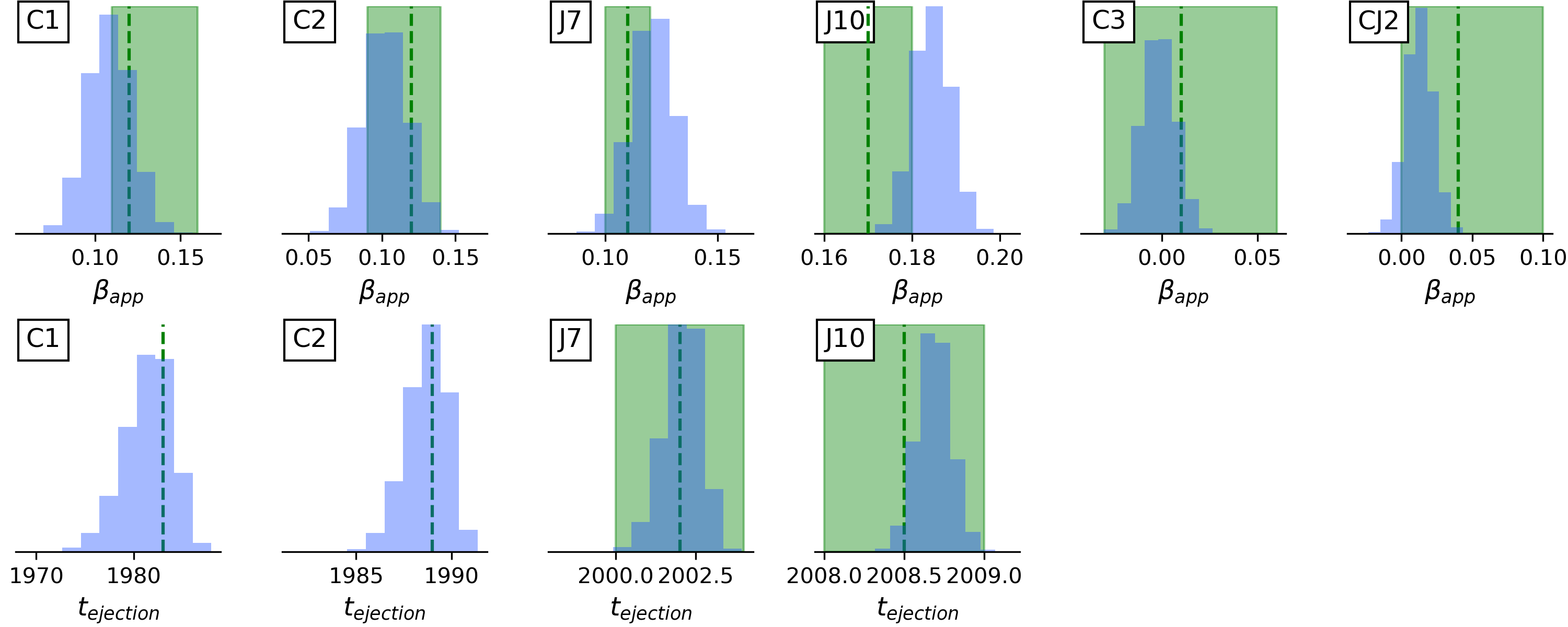}
\caption{The posterior of the fit apparent speed ($\beta_{app}$ is fractional light speed of the approaching jet component) and component ejection dates. The vertical green dashed line and the shaded area is the previously obtained value from the literature. Note that due to components C3 and CJ2 having very slow speeds, we do not obtain any significant ejection date (the posterior is distributed over a very large range of dates). }
\label{fig:compSpeeds}
\end{center}
\end{figure*}

For all the jet components, with the addition of our new epochs to the ones previously published in the literature, we fit for proper motion using a {\tt PYMC3} \citep{salvatier2016probabilistic} implementation of the Hamiltonian Monte-Carlo \cite[HMC][]{2011hmcm.book..113N} technique using a Gaussian likelihood. The fit proper motion along with $68\%$ high-density intervals are shown as a dashed black line and the filled gray area in Figure \ref{fig:properMotion}. We use flat priors for all the components but for J7 and J10 we use their previously determined values from TANAMI as priors\footnote{due to fewer number of data points for J7 and J10, our data was not informative enough to converge on a solution when using a uniform prior.}. The posterior of the component speeds and ejection dates from our are shown in Figure \ref{fig:compSpeeds} and are also provided in Table \ref{tab:kinematics}. All the model fit components used to study the jet kinematics are provided in Table \ref{tab:fitComps} of the APPENDIX section. In Figure \ref{fig:compSpeeds}, we find our measured apparent jet speeds and jet ejection dates to be within one sigma agreement with the values from the literature (shown as a green shaded area). \\

\subsection{SPECTRAL ANALYSIS} \label{sec:results:spectralAnalysis}
Due to the wide bandwidth of our VLBA observations ($256$\,MHz split into 8 IFs), we can have significant spectral averaging in regions with steep spectral indices when the full band is imaged. Hence, we mitigate this by imaging every IF independently, together giving us $12 \times 8$ IFs between $4.59 - 7.78$\, GHz for every epoch. We also flag the long baselines of higher frequency observations and the short baselines in the lower frequency observations in order to ensure all our individual IF images probe similar spatial scales of Centaurus A's jet. This is accomplished by using only the baselines between $2 - 36$ mega wavelengths and using a common restoring beam (BMAJ=$27$\,mas, BMIN=$9$\,mas, and BPA=$-14^{\circ}$) for all IFs and epochs. The individual IF images are inspected by eye, and images that show obvious artifacts such as significant negative structures are flagged. \\

\begin{figure}[h]
\begin{center}
\includegraphics[width=0.8\linewidth]{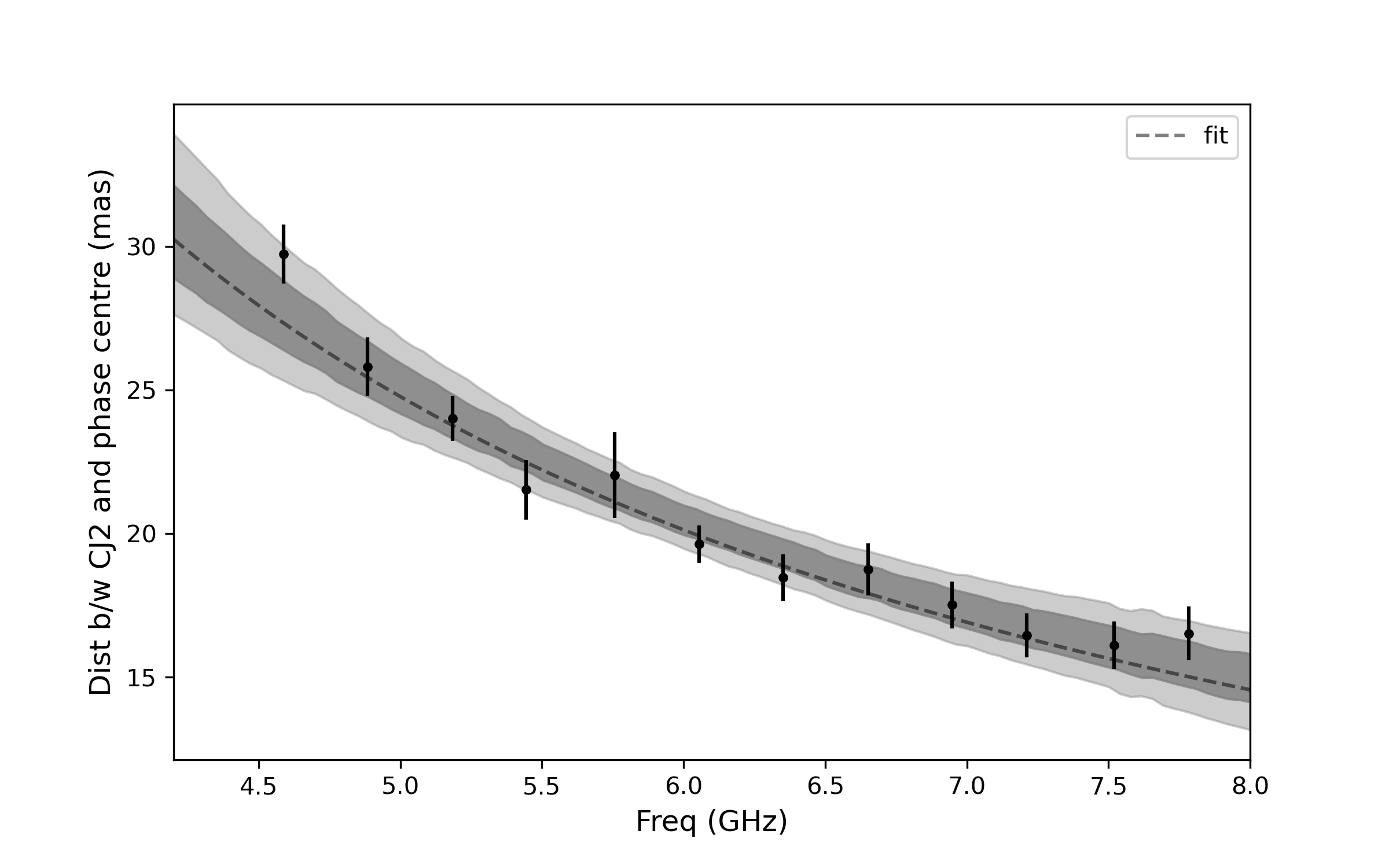}
\caption{The distance of the CJ2 component, as measured from the phase center (peak intensity location in the image) of the epoch-wise stacked image, with $95\%$ measurement errors. We also show the fitted core-shift function as the black dashed line, along with $68\%$ and $95\%$ high-density intervals of the fit in different shades of grey.}
\label{fig:coreshift}
\end{center}
\end{figure}

 As previously mentioned in Section \ref{sec:datareduction}, the absolute position of Centaurus A is lost due to having performed self-calibration that places the peak intensity region at the phase center.  For an expanding jet, the apparent peak intensity region moves downstream as $r\propto \nu^{-1/k}$ ($r$ is the distance from the black hole, $\nu$ is the observing frequency, and k is a constant that depends on the jet geometry), as longer wavelengths reach the optical depth $\tau \sim 1$ surface at larger distances from the base of the jet \citep{1979ApJ...232...34B}. This apparent shift downstream in the peak intensity region with frequency is referred to as the core-shift in literature \citep{sokolovsky2011vlba}. Hence our different IF images have undergone different offsets during self-calibration. We mitigate this unknown frequency-dependent offset between IFs by fitting for the theoretically motivated core-shift function to the CJ2's distance from the apparent jet base at a given frequency (i.e., the phase center), and we use the fit function to align the different frequencies as it always appeared as a discrete component with high signal to noise\footnote{Note that in later sections we find the CJ2 component to be free-free absorbed. In the presence of a large density gradient in the free-free absorbing screen, the apparent position of CJ2 is expected to shift with frequency. However, through manual inspection of the images at different frequencies, we find no evidence of systematic offsets between the frequencies when using CJ2 to lock reference frames, suggesting the free-free absorbing torus does not have a significant density gradient along the line of sight towards CJ2.}. Note that the approaching jet appears continuous once the above-mentioned $2-36$ mega wavelength cutoff is applied and hence image-based cross-correlation methods could not be employed for aligning the different IFs. In Figure \ref{fig:coreshift}, we show the epoch-averaged distance between the CJ2 component and the base of the jet (phase centre), along with the HMC fit of the physically motivated core-shift function. We obtain $k=0.9\pm0.1$ in the fit core-shift function and discuss its implications in Section \ref{sec:discussion:spectralAnalysis}.\\

\begin{figure*}[h]
\begin{center}
\includegraphics[width=\linewidth]{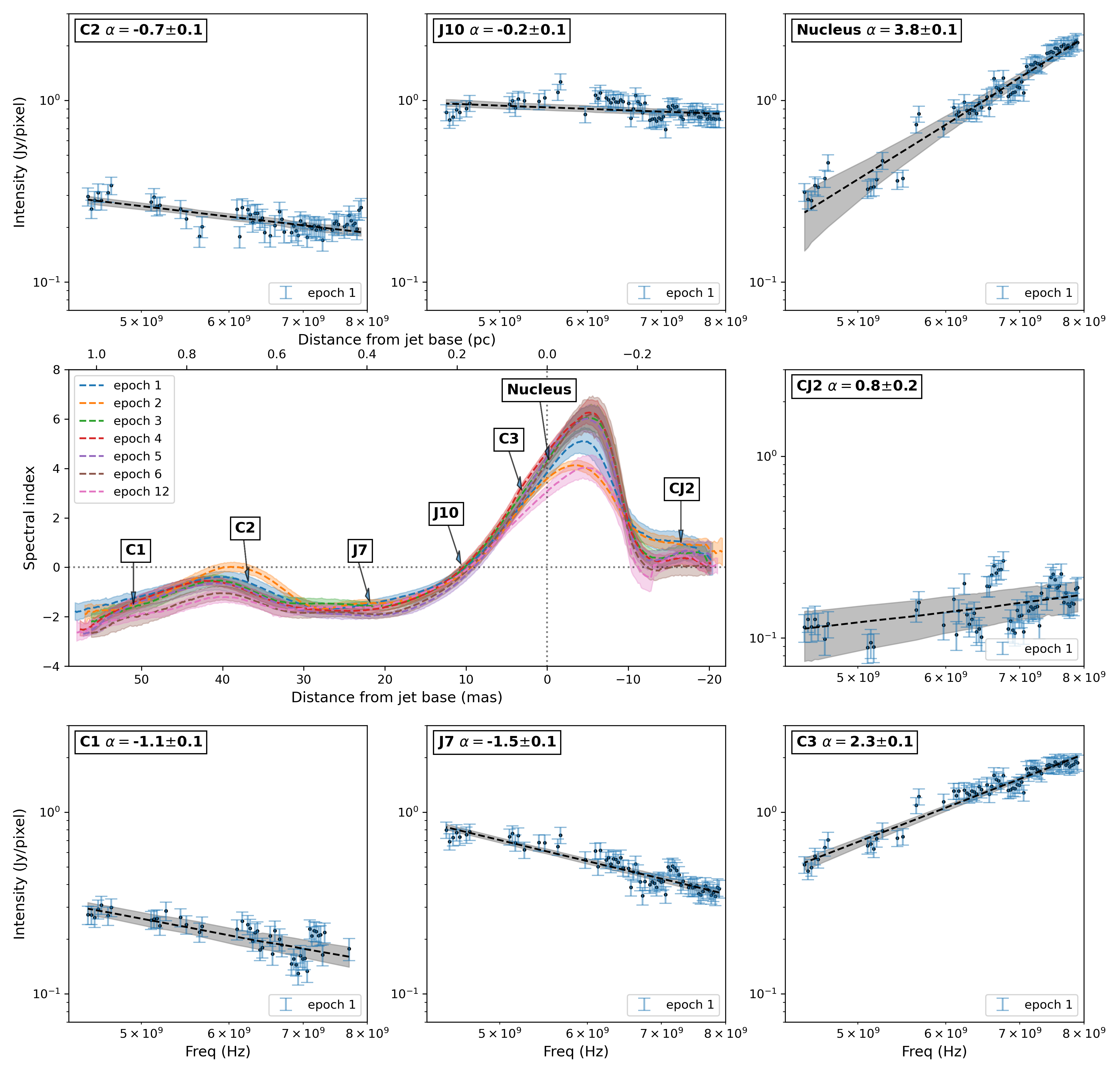}
\caption{The fitted spectral index (along with $68\%$ high-density intervals) along the ridge line of Centaurus A. Distances are measured from the base of the jet at $7.78$\, GHz, with positive distance values being along the approaching jet. In the zoom-in plots, we show the corresponding intensity values from each IF at the locations of jet components. Also, as can be seen from the central plot, we consistently get very similar spectral indices across all epochs.}
\label{fig:spectralIndex}
\end{center}
\end{figure*}

Having aligned the IFs, along the ridge line of Centaurus A (position angle $48^{\circ}$, measured East of North), using the IF images that had pixels above 10$\sigma$, we fit a power-law, and the resultant spectral indices (defined as $S_{\nu} \propto \nu^{\alpha}$, where $S_{\nu}$ is the intensity measured at a frequency $\nu$, and $\alpha$ is the spectral index) for different epochs are shown in the middle panel of Figure \ref{fig:spectralIndex}. As zoom-outs, we also show the spectrum and the fitted power-law at locations of interest (nucleus and jet components).\\

\subsection{LINEAR POLARISATION} \label{sec:results:linearPolarisation}
A montage of all linearly-polarised intensity maps of Centaurus A is given in APPENDIX \ref{sec:appendix:montage}. For regions with significant\footnote{regions with either Stokes Q or Stokes U over $4\sigma$ and simultaneously over $4\sigma$ in Stokes I image.} polarised emission, we show the polarised intensities along with the EVPAs of the electric field. We note from the montage that although the observed linear polarisation is very close to the detection limit, we see consistent regions of polarised emission (such as the leading edge of the approaching jet) across epochs and adjacent bands, thus building confidence in its significance.\\

The EVPAs of the observed polarised emission informs us of the direction of the local magnetic field responsible for the polarised emission. For optically thin regions of the jet, the EVPA should be perpendicular to the local magnetic field, and parallel for the optically thick parts of the jet \citep{o2009three}. However, as this emission passes through external cold plasma regions with magnetic fields along the line of sight, the intrinsic EVPAs ($\chi_{\circ}$) of the synchrotron emitting plasma undergo Faraday rotation and the observed EVPA ($\chi_{obs}$) can be mathematically represented as follows. 

\begin{equation}
\label{equation:faradayRotation}
    \chi_{obs}(\lambda) = \chi_{\circ} + RM.\lambda^{2},
\end{equation}
where RM is the rotation measure in $rad/m^{2}$, and $\lambda$ is the observed wavelength in m. The RM is an intrinsic property of the Faraday rotating material, calculated using the equation, 

\begin{equation}
    RM = 0.81 \int_{source}^{observer}n_{e}\vec{B}.\vec{dl},
\end{equation}
where $\vec{B}$ is the magnetic field in $\mu G$, $n_{e}$ is the electron density in units of $\rm{cm}^{-3}$, and $\vec{dl}$ is the path length along the line of sight measured in parsec. A positive RM would imply the magnetic field to have a component pointed towards the observer, and a negative RM would imply the magnetic field to point away from the observer. In principle, we can have multiple Farady rotating screens along the line of sight, the net effect of which could cause the observed low levels of linear polarisation in Centaurus A's jet. Alternatively, we could also have faint polarised emission in the individual IFs, but due to strong Faraday rotation (very high RM), the signal gets depolarised in the bandwidth-averaged images. In the next section, we investigate these possibilities by performing Rotation Measure (RM) synthesis across our entire frequency band. \\

\begin{figure*}[h]
\begin{center}
\includegraphics[width=0.9\linewidth]{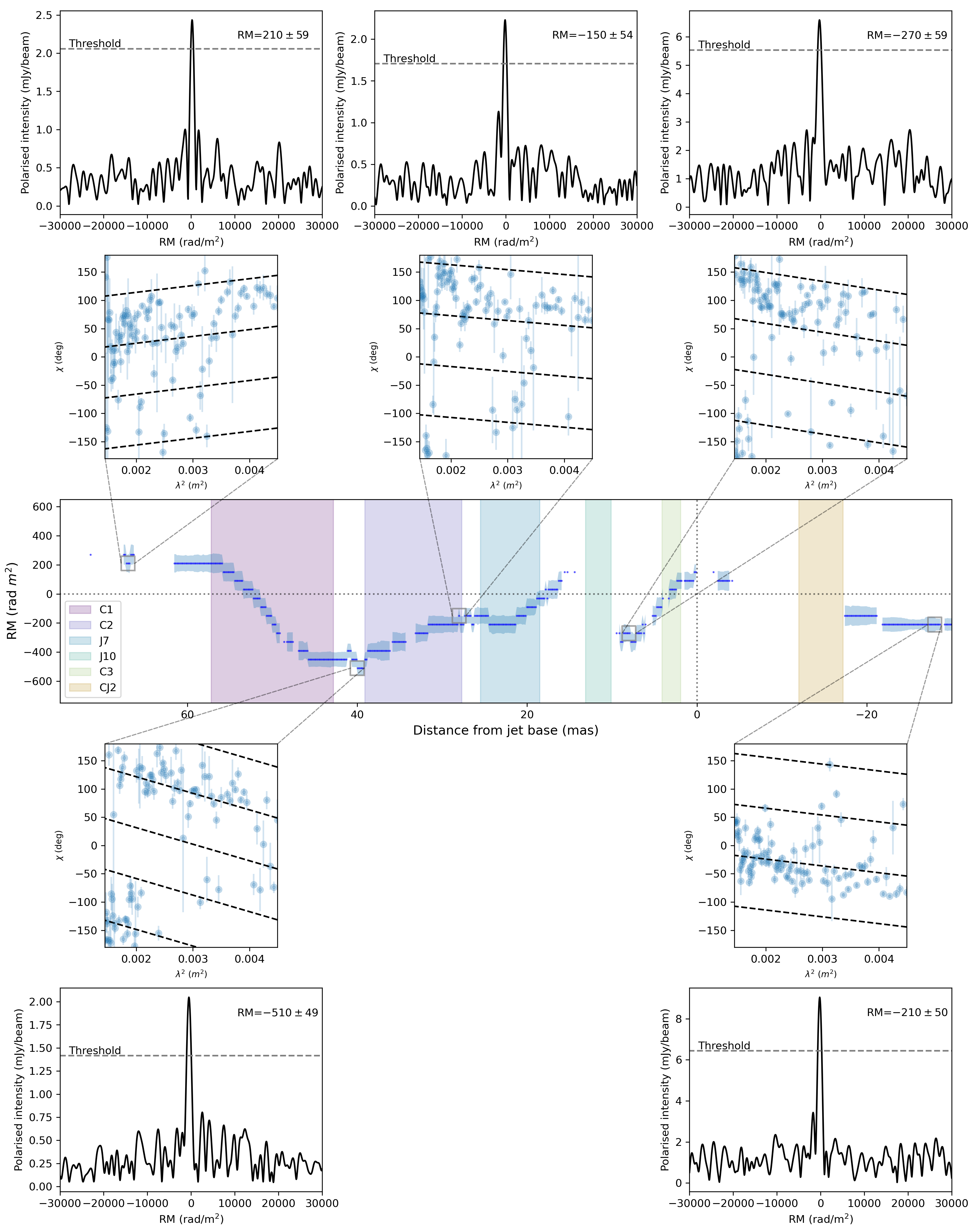}
\caption{The measured RM values with 1 sigma error along the ridge line of Centaurus A. To help better visualize the observed RM variations, the zoom-in plots show the observed EVPA at different IFs at different regions of inflection along the jet and the corresponding FDFs. The $8 \sigma$ noise threshold used in the FDF is shown using the dashed horizontal line.  The locations of jet components are shown as shaded regions (as obtained from our 7.8 GHz frequency). We note that due to having a larger restoring beam (due to having applied a constant uv range cutoff at all IFs), we obtain a larger beam in the images used for RM synthesis as compared to the beam size in the $7.8$\,GHz kinematics study, resulting in RM structures in the central panel spanning beyond C1 (CJ2) component along the approaching (receding) jet, due to image based beam convolution effects. The dashed black lines in the zoom-in panels help guide the reader to visualize the slope (Faraday depth) of the Faraday rotation material at a given location along the jet.}
\label{fig:rmSynth}
\end{center}
\end{figure*}

\subsubsection{Rotation Measure Synthesis} 
\label{sec:results:rmsynth}

RM synthesis \citep{brentjens2005faraday, heald2008faraday} uses Fourier transforms of the observed complex intensity (Q + iU, where Q is Stokes Q emission and U is Stokes U emission) at different frequencies to recover the different Faraday rotating screens at different Faraday depths ($\phi$). The output Fourier transform, $F(\phi)$, in units of $\rm{rad}/\rm{m}^2$, is called the Faraday dispersion function [FDF]. RM synthesis can also coherently integrate weak polarised emission across a wide bandwidth, hence enhancing its detection significance. From the observed polarised emission, $P(\lambda^2)$, we can recover the Faraday dispersion function using the equation (see \cite{brentjens2005faraday, heald2008faraday} for more details),

\begin{equation}
    F(\phi)=K \int_{-\infty}^{+\infty}P(\lambda^{2})e^{-2i\phi(\lambda^{2} - \lambda_{\circ}^{2})}d\lambda^{2},
\end{equation}
where $\lambda_{\circ}$ is an arbitrarily chosen reference frequency and K is the integral of the weights defined as $K = \int W(\lambda^2)$. Note that the the standard formalization of RM synthesis assumes constant polarised intensity at all frequencies which, given the broadband nature of our observation campaign, might produce spurious results. As described in \cite{brentjens2005faraday}, we mitigate this by using our spectral maps from Section \ref{sec:results:spectralAnalysis} to calculate the weights for the different IFs. \\

A comprehensive study of 149 active galactic nuclei by the MOJAVE survey \citep{hovatta2012mojave} found a range of rotation measures up to an absolute magnitude of $\approx 3,000$ $rad/m^{2}$. We take a more conservative approach and search all Faraday depths from $-30,000$ to $+30,000$ $rad/m^{2}$ in our RM-synthesis.  We apply uv-range cutoffs for all our bands such that they sample the same spatial scales and align them using our fit core-shift function, as done for our spectral maps. As we lack resolution in the transverse direction of the jet, we only perform RM-synthesis along the ridge line of the jet. Having first performed RM-synthesis for the individual epochs, we did not obtain sufficient SNR towards Centaurus A. Hence, we stack the individual EVPA-corrected complex intensity images (Q +iU) for every epoch, and then perform RM-synthesis on the time-averaged stack. Although performing a time-averaged analysis tends to wash out rapid variations that may be present in the observed sub-parsec scale jet, stacking has been shown to be an effective method for detecting the underlying long-term persistent magnetic field structures \citep{2023MNRAS.520.6053P}. \\

Regions along Centaurus A's ridge line that showed significant peaks in our RM-synthesis analysis (over $8\sigma$) are shown in Figure \ref{fig:rmSynth}. The SNR associated is calculated using the noise estimated near the wings of the real component of the FDF. As described in \cite{2020PASA...37...29R}, the uncertainty associated with the measured RM using RM-Synthesis depends on its SNR as described in Equation \ref{eq:rmsnr}. \\

\begin{equation}
    \delta RM  = \frac{\Delta (RM)}{2 \times SNR},
    \label{eq:rmsnr}
\end{equation}
where $\Delta (RM)=2\sqrt{3}/\Delta(\lambda^{2})$ is the RM resolution, and $\lambda^{2}$ is the wavelength squared measured across the entire observing bandwidth ($4.59-7.78$\,GHz). For the frequency range used in this study, the full-width half max of the Rotation Measure Spread Function (RMSF) is $\Delta (RM) = 1136$ rad/m$^2$. In the zoom-in panels of Figure \ref{fig:rmSynth}, we show the position angle measured in the individual IFs plotted as a function of wavelength squared and the corresponding FDF. Using the background black dashed lines, we help guide the reader to visualize the slope (Faraday depth of the rotating screen) that defines the linear relationship between the observed polarisation angle as a function of wavelength squared (Equation \ref{equation:faradayRotation}), where the slope of the dashed black line is obtained from our RM-synthesis. \\

\section{DISCUSSION} \label{sec:discussion}
\subsection{JET STRUCTURE AND KINEMATICS}
On the observed sub-parsec distances from the central black hole, the jet appears to have a collimated and mildly-relativistic outflow. We also obtain a reduced signal-to-noise detection of a receding jet due to Doppler de-boosting effects. Our kinematic results hint at low-inclination viewing angles of the jet, due to the observed significant differences in apparent speeds for components in the approaching jet and receding jet directions (discussed further in Section \ref{discussion:jetInclination}). In the approaching jet direction, we see a continuous jet, with distinct components moving downstream from the black hole. These components show a range of apparent speeds, with significant flux density variability in some components, hinting towards a faster evolving `inner jet' on spatial scales much smaller than probed here. In the lower-frequency bands ($<6.3$\,GHz images from the montage given in APPENDIX \ref{sec:appendix:montage}) we see a faint jet component about $90$\,mas from the nucleus. Although this component's kinematics were not discussed previously due to its optically thin nature (and hence its non-detection in the higher-frequency observations used in {\tt PAPERI/II}), the component has been previously detected in the $2.2$\,GHz and $5.0$\,GHz observations used in \cite{tingay2001estimates}. This is likely to be a component that was launched much before the C1 component, which has propagated downstream and adiabatically expanded beyond the spatial scales probed by the VLBA (hence the lower SNR). Given the range of approaching jet speeds observed in our kinematic analysis, we expect this downstream component to have been ejected between $1955-1981$, making it much older than all the other approaching jet components discussed here.  \\

Centaurus A's nucleus region shows a complex morphology, which always seemed best described using two elliptical Gaussians, with one of the components always being present  $\sim 4$\,mas downstream (0.07 pc). This nuclear downstream extension (C3 component), was first identified by {\tt PAPER I} as a slow-moving component during their observation campaign from 1988 - 1995. However, with the addition of a 5-year time baseline to already-published data, {\tt PAPER II} confirmed the structure to be a stationary feature in the jet with negligible proper motion. Using Space VLBI images made with sparse uv-coverage, \cite{fujisawa2000large} claimed this stationary C3 component to be a $52^{\circ}$ bend in the jet (and hence seen as a bright component due to changes in viewing angle), contradicting the higher resolution well-collimated jet observed at C3's location by TANAMI \citep{muller2014tanami} and \cite{horiuchi2006ten}. Furthermore, {\tt PAPER II} argues that due to the jet retaining a well-collimated structure downstream from the C3 component, it is unlikely that the jet undergoes a $52^{\circ}$ bend as suggested by \cite{fujisawa2000large}. We continue to see this C3 component to be a stationary component during our observation campaign. As the component shows low levels of linear polarisation (in some of the frequencies and epochs as can be seen in the montage provided in APPENDIX \ref{sec:appendix:montage}), we suspect it to be a re-collimation shock, as expected from magnetohydrodynamic simulations of AGN jets.  Similar stationary re-collimation shocks are observed in M87 at subparsec distances from the central SMBH \citep{2023arXiv230711660N, walker2018structure}.  \\

Another point of interest in the approaching jet is the J7 component, about $22$\,mas downstream from the nucleus. As can be seen from Figure \ref{fig:lightCurve}, the component shows rapid flux density variations within times scales of a few weeks. During 1991-1992, when the C1 component spatially coincided with the current location of J7, {\tt PAPER I} observed C1 to show rapid variations in flux density (as can be seen in Figure \ref{fig:lightCurve}). {\tt PAPER II} again found similar flux density variation in the C2 component during 2001, when it traveled past J7's current location in the jet, all of which hinted towards the rapid evolution of jet components at a fixed distance from the nucleus. The higher resolution TANAMI images of Centaurus A identified a jet widening `tuning fork' feature at this location, that they found to be best explained as an interaction of jet with a star (see \cite{muller2014tanami} for the different possible scenarios explored to explain the jet widening), and we continue to see this location to remain a region of rapid jet evolution in our observations (see Figure \ref{fig:lightCurve}). \\

Under the premise that the rapid variability at J7's current location is caused by a star crossing the jet stream, \cite{muller2014tanami} predicts the jet crossing time for the star to be about 20 years. The VLBA observations used in this work are exactly 20 years since the earliest detection of rapid variability by {\tt PAPER I} in the early 1990s. However, in order to confirm the hypothesis, we need future VLBI observations to show no variability at the point of interest as we expect the star to have crossed the jet. Assuming a circular Keplerian orbit for the star around the central black hole, the 20 years of observed jet crossing time would imply a jet opening angle of $\le 0.6 ^{\circ}$ (see APPENDIX \ref{sec:appendix:openingAngle}). However, as this is a strongly model-dependent calculation, we resort to using the conventional method of using the lack of resolved structures downstream (perpendicular to the jet) to place an upper limit on the jet half-cone opening angle to be $\le 17^{\circ}$, when projected along the plane of the sky. If the inclination of the jet were to be $\approx 30^{\circ}$ (further discussed in Section \ref{discussion:jetInclination}), this limit would reduce by a factor of 2, i.e, $\le 9^{\circ}$.\\

\subsection{SPECTRAL ANALYSIS} \label{sec:discussion:spectralAnalysis}
From Figure \ref{fig:spectralIndex}, we see that consistently for all epochs, the compact nucleus displays a highly inverted spectrum, which becomes optically thin for distances greater than $10$\,mas downstream. As previously observed by \cite{tingay2001estimates}, the nucleus of Centaurus A has a higher spectral index than permitted by synchrotron self-absorption alone. This is thought to be due to a free-free absorbing torus, through which we view the central engine of Centaurus A \citep{meisenheimer2007resolving}. A low-inclination viewing angle along with a free-free absorbing torus also best explains the inverted spectrum emission from the CJ2 receding jet component (discussed further in Section \ref{sec:discussion:natureOfCJ2}). \\

Downstream along the approaching jet, the jet becomes optically thin, often displaying a spectral indices as steep as $\alpha=-2$. Such steep spectral indices have been observed in other radio galaxies, such as M87, and have been used alongside jet models to infer the injection rate of non-thermal electrons into the jet \citep{2023A&A...673A.159R}. In future, our measurement of spectral distribution along the jet can be used in conjunction with comprehensive jet models to infer similar properties about the jet in Centaurus A. Further downstream, the spectrum flattens (becomes optically thicker, compared to emission between $15-30$\,mas) for distances greater than $35$\,mas. Spectral flattening is often interpreted as regions of particle acceleration, such as from shocks traveling along the jet \citep{2014AJ....147..143H}, and is consistent with the location of C2 component. Coincidentally, this region also shows a rapid change in RM, possibly hinting at the possibility of the jet being deflected by a clump of gas (see Section \ref{sec:discussion:polarisationAnalysis}), thus resulting in the spectrum flattening due to density enhancements. \\ 


\subsection{POLARISATION ANALYSIS} \label{sec:discussion:polarisationAnalysis}
VLBI polarisation experiments of AGN \citep[see e.g.,][]{hovatta2012mojave, lister2018mojave, pushkarev2023mojave} have helped understand jet collimation and acceleration very close to the central black hole by probing the geometry of the magnetic field responsible for the relativistic jets. However, for torus-obscured radio galaxies, the observed polarisation levels are often very low due to foreground absorption and external depolarisation. From our EVPA-corrected linear polarisation images of Centaurus A (APPENDIX \ref{sec:appendix:montage}), we find very low levels of linear polarisation spread across different regions of the jet at different epochs and frequencies. However, we note that the polarisation feature at the leading edge of the jet ($\approx 50$\,mas from core) is consistent across multiple epochs for frequencies below $4.9$\,GHz, hence enhancing its significance. \\

For frequencies between $5.7 - 6.6$\,GHz, we see linearly polarised emission near the nucleus. Being able to observe polarisation near the nucleus would also suggest the free-free absorbing torus to be patchy in nature, because otherwise, it is unlikely for the linear polarisation to survive propagation through such a dense region. The observed EVPAs near the nucleus region vary significantly between epochs. This could be due to sampling different parts of complex un-resolved jet structure due to changing uv-coverage between observations. Alternatively, this could also be explained as due to the ejection of new components at the nucleus causing changing EVPA as the ejecta traverse varying magnetic fields. If the observed linear polarised emission is from the C3 component, then it would suggest the stationary feature to be a standing re-collimation shock at subparsec distances from the central black hole in Centaurus A.\\

Our RM-synthesis analysis performed on the stacked VLBA image shows the presence of Faraday rotating plasma along the line of sight towards the source (Figure \ref{fig:rmSynth}). The measured RM values are much higher than the $\approx -52.9$\, rad/m$^2$ Milky Way Galactic RM contribution towards Centaurus A \citep{2009ApJ...707..114F, 2013ApJ...764..162O}, suggesting the presence of other Faraday rotation screens closer to the jet. The fluctuations in RM values at mas scales seen in Figure \ref{fig:rmSynth} could be explained as viewing the jet through clumpy gas surrounding the jet. The clumpy nature of the torus is further supported by the fact that we observe significant polarisation from the CJ2 receding jet component, which should otherwise be expected to be absorbed for a low-inclination angle AGN (inclination angle of the jet is later constrained to be $\le 25^{\circ}$ in Section \ref{discussion:jetInclination}). The rapid change in $\|RM\|$ at $\approx 40$ mas could be interpreted as the jet being deflected by a clump of gas, as previously observed in 3C120 \citep{2008MmSAI..79.1157G}. Alternatively, this change in RM sign can also be interpreted as the change in the dominant line of sight magnetic component due to accelerating/decelerating flows \citep{o2009three}. The proposed onset of acceleration towards the leading edge of the jet is also consistent with the astrometrically observed onset of acceleration in the higher-resolution TANAMI images for components $>40$\,mas from the nucleus. Moreover for Centaurus A, the higher jet speeds seen downstream at kilo-parsec distances from the central engine \citep{2003ApJ...593..169H}, are best explained through the acceleration of the slower sub-parsec jet components observed in our VLBA campaign.\\

\begin{figure*}[h]
\begin{center}
\includegraphics[width=0.5\linewidth]{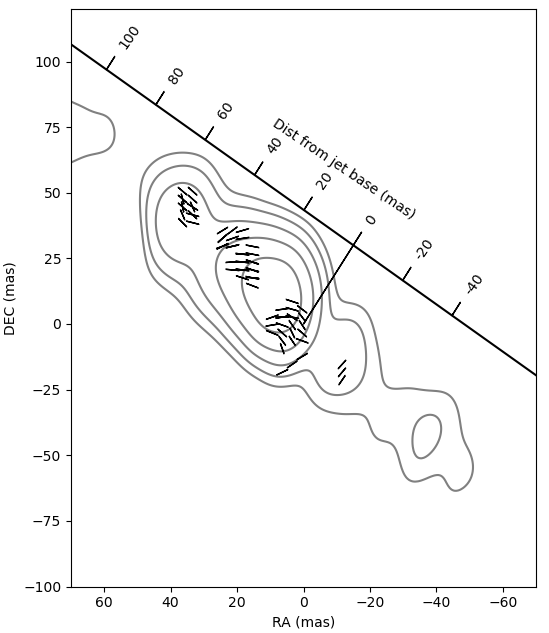}
\caption{The Faraday rotation corrected intrinsic electric field directions in the subparsec scale jet of Centaurus A. In the background, we show the stacked  $4.59$\, GHz image having applied the uv-range cutoff described in Section \ref{sec:results:spectralAnalysis}. As the diagonal axis, we show the angular distance downstream as measured from the jet base at $7.8$\, GHz. As the electric fields are measured from the epoch-stacked image, they represent the directions of long-term persistent fields in the jet. }
\label{fig:globalFieldDirection}
\end{center}
\end{figure*}

The epoch-stacked linear polarization images from every band are shown in APPENDIX \ref{sec:appendix:montageStacked}, and the corresponding Faraday rotation corrected intrinsic electric field directions are shown in Figure \ref{fig:globalFieldDirection}. We find the presence of ordered magnetic fields but with changes in field directions downstream along the jet. In the optically thin leading edge of the jet, the majority of the electric fields are parallel to the jet, suggesting the presence of a helical/toroidal magnetic field. Near the nucleus, we see rapid changes in field direction suggesting either the presence of variable magnetic fields close to the central black hole or ordered magnetic fields that are changing on scales smaller than the beam. The $\sim 90^{\circ}$ difference in the jet EVPAs at 50 mas downstream the approaching jet and the CJ2 receding jet component possibly suggests evidence for a stratified jet structure where the magnetic field could be toroidal near the spine (dominating the emission in the approaching jet). With the de-boosted spine of the receding jet, we mainly see a shear layer of longitudinal field. However, other combinations of geometrical orientations of magnetic fields, optical depth effects, and beam effects could also explain the $90^\circ$ flip in EVPA. \\



 \subsection{JET INCLINATION}
\label{discussion:jetInclination}
Depending on the method used, a tension exists in the literature between the different estimates of Centaurus A's inclination angle at sub-parsec distances from the black hole. The difference between the apparent proper motion of the C1/C2 approaching jet components and the apparent motion of the CJ2 receding jet component constrains the inclination angle to be between $50^{\circ}-80^{\circ}$ ({\tt PAPER II}), whilst the brightness ratio measurements of approaching jet and receding jet components has yielded an inclination angle of $12^{\circ}-45^{\circ}$ (TANAMI, \cite{muller2014tanami}). We attribute this tension to two key identified caveats: firstly, as will be discussed in further detail in Section \ref{sec:discussion:natureOfCJ2}, CJ2 is not the receding jet equivalent of either the C1 or C2 approaching jet components. Second, having identified the receding jet to be free-free absorbed in Section \ref{sec:results:spectralAnalysis}, the measured brightness ratio could underestimate the true brightness ratio of the underlying jet. Hence, we re-estimate the inclination, avoiding the use of the semi-stationary optically thick CJ2 receding jet component that we suspect to be the cause of the tension. \\

\begin{figure*}[h]
\begin{center}
\includegraphics[width=0.8\linewidth]{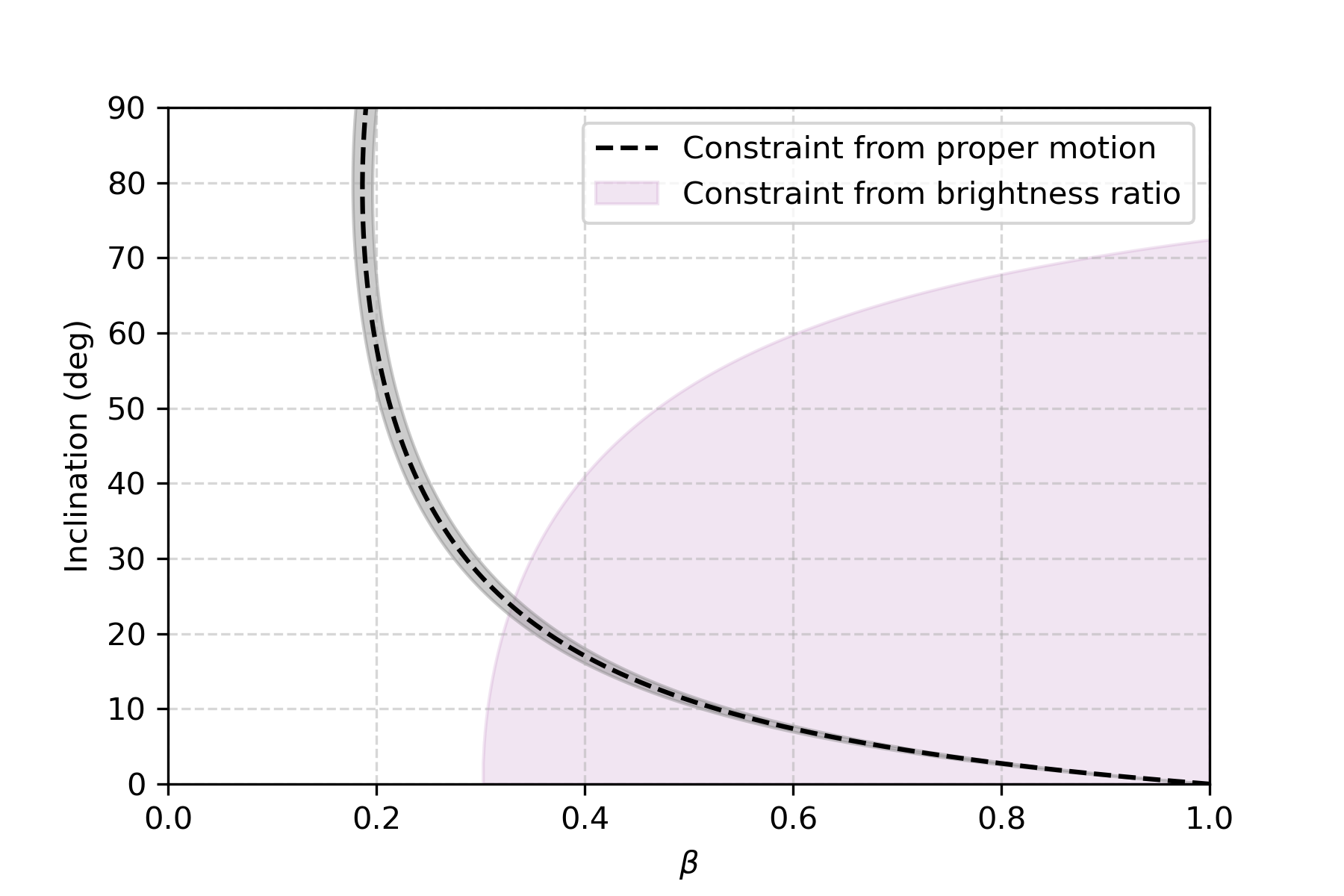}
\caption{The constraints on inclination angle and intrinsic jet speed obtained from the proper motion and brightness ratio of Centaurus A's approaching jet components. Together they constrain the inclination angle of Centaurus A to be less than $25^{\circ}$.}
\label{fig:inclination}
\end{center}
\end{figure*}

Using the apparent speed $\beta_{app}$ of the fastest approaching jet component (J10), we can obtain the combination of possible intrinsic jet speeds ($\beta$) and inclination angle ($\theta$), using Equation \ref{Eq:inclinationConstrain1}. This is also shown as the grey shaded area in Figure \ref{fig:inclination}. \\

\begin{equation}
    \beta = \frac{\beta_{app}}{\sin(\theta) + \beta_{app}\cos(\theta)}
    \label{Eq:inclinationConstrain1}
\end{equation}

The lack of a detected receding jet equivalent for approaching jet components can be used to place a lower limit on the true brightness ratio ($r$). This lower limit on the brightness ratio further informs us of the possible combinations of $\theta$ and $\beta$, using Equation \ref{Eq:inclinationConstrain2}. \\

\begin{equation}
    \theta < \cos^{-1} \left(\frac{1}{\beta} \times \frac{r^{\frac{1}{k-\alpha}} - 1 }{r^{\frac{1}{k-\alpha}} + 1}\right),
    \label{Eq:inclinationConstrain2}
\end{equation}
where $k=3$ for discrete jet components and $\alpha$ is the spectral index of the jet component. To mitigate the influence of the free-free absorbing torus, we use C1 (or C2) component's brightness during our 2013 epoch to determine a lower limit on the jet's brightness ratio\footnote{The two components have traveled far enough downstream for them not to be influenced by the torus. This is further supported by their optically thin spectrum seen in Section \ref{sec:results:spectralAnalysis}}. The ratio of the C1 (or C2) component brightness to the occasionally detected CJ1* component's brightness (instead of the image noise) can be used to obtain a conservative estimate of $r>=12$ as the lower limit on the brightness ratio, and the corresponding constraints on $\theta$ and $\beta$ are shown in Figure \ref{fig:inclination}. \\

Using the above two equations, we constrain the inclination angle of Centaurus A's sub-parsec jet to be $<25^{\circ}$, in agreement with the $\approx 16^{\circ}$ inclination determined for the kilo-parsec scale jet seen by the VLA \citep{2003ApJ...593..169H}, and the recent $\le 41^{\circ}$ constraint from X-ray observations \citep{bogensberger2024superluminalpropermotionxray}. This also places a lower limit of $\beta>=0.33$\,c for the fast-moving J10 component, and $0.19$\,c as the lower limit for the remainder slower approaching jet components. If the jet inclination was the same at the sub-parsec distances from the central black hole as the $16^{\circ}$ inclination measured for the kilo-parsec scale jet, this would make J10's jet speed to be $0.41$\,c and $>=0.27$\,c as the lower limit for the other slower components. As the jet observed very close to the black hole by the Event Horizon Telescope \citep{janssen2021event}, the subparsec scale jet seen by the VLBA in this work, and the kilo-parsec scale jet seen by the VLA \citep{2003ApJ...593..169H}, all seem to have a similar position angle along the sky (see Figure \ref{fig:zoomin}), it is reasonable to assume that they all have similar inclination angles, as it is less likely for a jet to only bend along the line of sight, with no changes in position angle in the plane of the sky. However, at 10-100 kiloparsec distances from the black hole, \cite{2022NatAs...6..109M} measures Centaurus A's inclination angle to be $\approx 68^{\circ}$ suggesting a bend in the jet at the large scales made using MWA \citep{tingay2013murchison}, which is further supported by the change in the position angle of the large scale jet when projected on the plane of the sky, as shown in Figure \ref{fig:zoomin}.\\


\subsection{JET EXPANSION PROFILE}
The value of $k$ in the fit core-shift function (in Section \ref{sec:results:spectralAnalysis}) informs us of the jet expansion geometry \citep{2024MNRAS.528.2523N} near the base of the jet. Multiple studies of AGN \citep[to name a few]{2017MNRAS.468.4992P, 2012ApJ...745L..28A, nakamura2013parabolic} have found $k \approx 0.5$ for frequencies that probe very close to the black hole, suggesting a parabolic expansion profile due to the plasma still being accelerated. Further downstream, at $10^{5} - 10^{9}$ $R_{g}$ (the sub-parsec distances probed in this work) from the central black hole  \citep{spruit1997collimation, komissarov2007magnetic, lyubarsky2009asymptotic, boccardi2017radio}, a constant bulk speed is achieved, resulting in a conical jet expansion geometry, as proposed by \cite{1979ApJ...232...34B}, resulting in a value of $k=1$. Our measured $k=0.9\pm0.1$ in Centaurus A is in agreement with the trend observed in other AGN, and suggests a conical jet expansion geometry with a constant bulk speed for the regions of the jet traced by the optical depth $\tau \sim 1$ surface at $4.59 - 7.78$\, GHz frequencies, i.e., within 0.3 pc from the nucleus. Following the analysis done by \cite{2012Natur.489..326H} for M87, based on the asymptotic value of the core-shift function at higher frequencies, we expect the black hole in Centaurus A to be at least $10$\,mas upstream from the observed base of the jet at $7.8$\,GHz (assuming the jet continues being conical upstream). However, as the Event Horizon Telescope (EHT) finds evidence of a changing opening angle (i.e, quasi-parabolic geometry) near Centaurus A's black hole \citep{janssen2021event}, the location of the black hole is expected to be at a much lower distance than the $10$\,mas asymptotic core-shift estimate. \\

\subsection{NATURE OF CJ2 COMPONENT}
\label{sec:discussion:natureOfCJ2}
The CJ2 component along the receding jet has always shown complex behavior at VLBA sensitivities. Since its earliest detection in 1992, the component has consistently been observed at all epochs\footnote{except a few observations in {\tt PAPER I} due to increased noise levels in the images.} with high signal-to-noise, about $15$\,mas southeast of the nucleus. During the 21 years of its detection (1992-2013), the component has shown no significant displacement ($<5$\,mas), and an apparent velocity ($0.01 \pm 0.01$\,c) that is consistent with it being a stationary component. Our spectral analysis of the feature (Figure \ref{fig:spectralIndex}) has shown CJ2 to be inverted spectrum  ($\alpha >= 0$), thus warranting further questions about its nature, such as, is CJ2 the receding component associated with the approaching jet components C1/C2? Or is CJ2 the optical depth $\tau \sim 1$ surface along the receding jet? Or is CJ2 a stationary nuclear extension along the receding jet? We discuss these scenarios below. \\

Since their estimated ejection times, C1 and C2 have traveled about 40-50 mas downstream along the approaching jet. If the CJ2 component were to be the associated receding jet feature of either C1 or C2, we would require inclination angles $<2^{\circ}$ to accommodate the very small apparent speed of CJ2 and its negligible proper motion during the 20 years (see APPENDIX \ref{sec:appendix:natureOfCJ2}). We find such small viewing angles to be unlikely for Centaurus A due to the observed free-free absorption towards the nucleus that requires larger inclination angles in order to view the nuclear region of the jet through the free-free absorbing external material. \\


From the spectral analysis of Centaurus A's jet (Section \ref{sec:discussion:spectralAnalysis}), we find evidence for the presence of free-free absorbing material surrounding the compact nucleus (i.e, the base of the approaching jet). For this reason, if CJ2 were to be the optical depth $\tau \sim 1$ surface along the receding jet, we would expect it to show a more inverted spectrum than the base of the jet along the approaching jet due to viewing it through a larger column depth of the free-free absorbing material. This is in contrast with what we observe in our spectral index measurement for CJ2, thus ruling out the possibility of it being the optical depth $\tau \sim 1$ surface along the receding jet. \\

It is plausible for CJ2 to be a semi-stationary feature along the receding jet, similar to the C3 jet component. A low inclination viewing angle (discussed further in Section \ref{discussion:jetInclination}) would require us to view this semi-stationary feature through the free-free absorbing materials surrounding the base of the jet, thus making it inverted spectrum (in agreement with the approx. $\alpha=1$ observed in Figure \ref{fig:spectralIndex}). Much like the C3 component, CJ2 also shows low levels of linear polarisation at some epochs/frequencies (see APPENDIX \ref{sec:appendix:montage}). The higher resolution TANAMI observations of Centaurus A show multiple smaller jet components that spatially coincide with our CJ2 component. These smaller components are very close to the detection threshold of the TANAMI data and are not detected across multiple epochs, and thus their kinematics were not investigated \citep{muller2014tanami}. Hence, it is also possible that our resolved CJ2 component is the emission centroid of multiple small receding jet components near the base of the jet. As jet components travel downstream, they adiabatically expand causing them to exponentially fade with distance. For a receding jet this ``fading'' happens at much shorter distances due to Doppler deboosting effects. Hence, it is possible that the emission centroid of a constant supply of receding jet components that rapidly fade past the detection limit can appear as a persistent resolved semi-stationary feature, much like the CJ2 component. \\

\section{CONCLUSION} \label{sec:conclusion}
In this study, we present the results from the first spectral and polarimetric VLBI study of the sub-parsec jet outflows in the Centaurus A radio galaxy, gaining insights into the jet structure and evolution at sub-parsec distances from the central supermassive black hole. We constrain many key jet parameters, such as the kinematics, inclination angle, opening angle, and expansion profile. We place a robust upper limit of $25^{\circ}$ on the inclination angle, which in turn provides key constraints on the intrinsic speed of the jet. Using core shift measurements, we find the jet to have a conical outflow with a constant velocity near the base of the jet. However, towards the leaouding edge of the Centaurus A's jet, through polarimetric analysis we find evidence to suggest the onset of acceleration.  These findings collectively contribute to our overall understanding of jet formation, collimation, and evolution in Centaurus A, and therefore AGN in general.\\

\section{Software and third party data repository citations} \label{sec:cite}
We acknowledge the work and the support of the developers of the following Python packages:
Astropy \citep{theastropycollaboration_astropy_2013,astropycollaboration_astropy_2018, The_Astropy_Collaboration_2022}, Numpy \citep{vanderwalt_numpy_2011}, Scipy  \citep{jones_scipy_2001}, matplotlib \citep{Hunter:2007}, SkyField\footnote{\url{https://rhodesmill.org/skyfield/}}, and PYMC3 \citep{salvatier2016probabilistic}. \\

The code required to reproduce all the figures in the paper can be obtained from our GitHub\footnote{\url{https://github.com/StevePrabu/Centaurus-A-VLBA-BO043}} and Zenodo \citep{prabu_2024_13999389} repositories. The {FITS} files of the images made in this study are available upon request. \\

\begin{acknowledgments}
The authors would like to thank Michael Janssen and Martin Hardcastle for kindly sharing the EHT and VLA images of Centaurus A. SPO acknowledges support from the Comunidad de Madrid Atracción de Talento program via grant 2022-T1/TIC-23797, and grant PID2023-146372OB-I00 funded by MICIU/AEI/10.13039/501100011033 and by ERDF, EU.
\end{acknowledgments}

%

\vspace{5mm}





\appendix




\begin{table*}[h!]
    \caption{Best fit {\tt difmap} models for the $7.8$\,GHz data. S is the model fit flux density of the component, d is the distance of the component from the phase centre, $\theta$ is the position angle of the component, A is the major axis of the component, B/A is the ratio of the minor and major axis (for an elliptical Gaussian), and $\phi$ is the position angle of the major axis of the component. }
    \centering
    \label{tab:fitComps}
    \begin{tabular}{@{}lcccccc@{}}
    S & d & $\theta$ & A & B/A & $\phi$  & ID \\
    (Jy) & (mas) & (deg) & (mas) & & (deg) & \\
    \hline \hline    
    \multicolumn{1}{c}{01/27/2013 (BO043A) } \\
    \hline 
    1.13416 & 0.764768 & -95.0850 & 3.54447 & 0.00000 & 37.2908 & core \\ 
    0.0849613 & 51.2439 & 50.3393 & 7.86070 & 0.00000 & -40.8916v& C1 \\
    0.584269 & 36.0836 & 48.6784 & 21.3402 & 0.144904 & 52.0457 &  C2\\
    0.273595 & 20.3077 & 46.4530 & 5.81051 & 0.718412 & 63.3824 & J7 \\
    0.671515 & 10.6301 & 47.6092 & 7.29614 & 0.344856 & 55.3234 & J10 \\
    1.12247 & 1.61428 & 52.9279 & 6.67498 & 0.356308 & 34.6580 & C3 \\
    0.0359646 & 39.4471 & -133.316 & 0.00000 &1.00000 & 0.00000 & CJ1$^{*}$ \\ 
    0.266711 & 15.3796 & -133.760 & 12.8848 & 0.525441 & 29.1303 & CJ2 \\
    \hline 
    \multicolumn{1}{c}{02/11/2013  (BO043B) } \\
    \hline
    0.268649 & 0.923460 & -172.426 & 4.06608 & 0.00000 & 6.29507 & core \\
    0.206292 & 49.9483 & 49.5416 & 11.8550 & 0.636584 & 37.6030 & C1 \\
    0.485724 & 34.3634 & 48.0527 & 14.6582 & 0.309046 & 61.0616 & C2 \\
    0.265257 & 20.5065 & 50.2928 & 7.34712 & 0.00000 & -30.7047 & J7\\
    0.614267 & 11.4519 & 49.4838 & 4.02266 & 8.05075e-09 &  74.1759 & J10 \\
    2.21849 &  1.02334 & 2.32560 & 6.78369 & 0.248519 & 48.0473v  & C3 \\
    0.0553808 & 50.7309 & -143.838 & 0.00000 & 1.00000 & 0.00000 & CJ1$^{*}$ \\
    0.255186 & 15.5762 & -127.201 & 10.9094 & 0.00000 & -61.2895 & CJ2 \\
    \hline
    \multicolumn{1}{c}{03/06/2013  (BO043C) } \\
    \hline
    1.61154 & 0.758030 & -90.9248 & 4.55352 & 0.268286 & 30.5919 &core \\
    0.323715 & 45.9212 & 49.5660 & 24.8243 & 0.00000 & 45.6901 & C1 \\
    0.192939 & 34.0257 & 48.6573 & 16.2363 & 7.97856e-09 & 44.5693 & C2 \\
    0.302464 & 21.2130 & 49.3513 & 8.53706 & 0.0616162 & 52.9576 & J7 \\
    0.698014 & 9.78115 & 50.7509 & 8.98535 & 0.347840 & 37.2911 & J10 \\
    0.779665 & 2.37614 & 54.5272 & 5.09232 & 0.267362 & 32.1924 & C3 \\
    0.172923 & 15.3328 & -126.459 & 6.60442 & 0.00000 & 72.8107 & CJ2 \\
    \hline
    \multicolumn{1}{c}{ 04/06/2013  (BO043D) } \\
    \hline
    1.46516 & 1.41557 & -139.441 & 2.86099 & 0.384725 & 40.7781 & core \\
    0.229235 & 47.9213 & 49.7710 & 16.4299 & 0.312466 & 39.1063 & C1 \\
    0.669907 & 25.6588 & 49.8453 & 25.6147 & 0.0962005 & 50.6893 & C2 \\
    0.0695673 & 20.6264 & 50.0964 & 6.97285 & 0.00000 & -3.94905 & J7 \\
    0.544596 & 10.4502 & 50.0213 & 8.82214 & 0.417642 & 30.2899 & J10 \\
    1.13369 & 2.16590 & 53.0629 & 5.00532 & 0.533699 & 30.6709 & C3 \\
    0.141951 & 14.6999 & -125.559 & 9.13123 & 0.117925 & 41.2231 & CJ2\\
    \hline
    \multicolumn{1}{c}{ 04/23/2013  (BO043E) } \\
    \hline
    0.899543 & 2.38666 & -169.232 & 2.72122 & 0.526081 & -6.38224 & core\\
    0.264144 & 48.8214 & 50.1487 & 16.0998 & 0.00000 & 40.0383 & C1\\
    0.387307 & 33.8979 & 49.1186 & 18.6896 & 0.460439 & 18.2125 & C2\\
    0.240399 & 21.0981 & 51.5031 & 9.42553 & 0.388243 & 14.8606 & J7\\
    0.588852 & 10.6407 & 55.2442 & 7.55676 & 0.00000 & 52.5295 & J10\\
    1.55395 & 1.65125 & 70.6115 & 4.80032 & 0.855508 & 21.3381 & C3\\
    0.0359845 & 40.4946 & -128.562 & 0.00000 & 1.00000 & 0.00000 & CJ1$^{*}$\\
    0.176982 & 16.6928 & -131.362 & 6.11435 & 0.461710 & 60.6397 & CJ2 \\
    \hline
    \multicolumn{1}{c}{ 05/11/2013   (BO043F) } \\
    \hline
    1.69199 & 0.934970 & -122.547 & 3.78146 & 0.320002 & 33.4070 & core\\
    0.213078 & 49.3625 & 49.0063 & 26.4556 & 0.0470682 & 38.3097 & C1\\
    \hline
    \multicolumn{1}{c}{Continued on next page...} \\
\end{tabular}\label{blah}
\end{table*}

\addtocounter{table}{-1}
\begin{table*}[h!]
    \caption{...continued from previous page.}
    \centering
    \begin{tabular}{@{}lcccccc@{}}
    S & d & $\theta$ & A & B/A & $\phi$  & ID \\
    (Jy) & (mas) & (deg) & (mas) & & (deg) & \\
    \hline
    0.199217 & 29.3059 & 52.8659 & 16.9363 & 0.00000 & 41.0155 & C2\\
    0.177265 & 21.9134 & 49.6835 & 5.78086 & 0.00000 & 40.0827 & J7\\
    0.671656 & 9.97357 & 51.9080 & 9.14439 & 0.333274 & 36.1777 & J10\\
    0.796058 & 2.69608 & 52.5278 & 3.83195 & 0.439515 & 23.4510 & C3\\
    0.142039 & 15.6188 & -123.918 & 7.33449 & 0.323644 & 46.5601 & CJ2\\
     \hline    
    \multicolumn{1}{c}{ 07/06/2013   (BO043I) } \\
    \hline
    1.92410 & 0.968099 & -115.461 & 4.92705 & 0.372950 & 33.2227 & core\\
    0.0409272 & 37.5834 & 50.1191 & 0.00000 & 1.00000 & 0.00000 & C2\\
    0.277874 & 20.8687 & 56.0815 & 16.4838 & 0.284990 & 17.4258 & J7\\
    0.681915 & 10.3986 & 48.4667 & 7.97737 & 0.392527 & 22.3492 & J10\\
    0.716639 & 3.08422 & 47.6363 & 6.45694 & 0.111506 & 20.8346 & C3\\
    0.131362 & 15.3195 & -125.181 & 11.8796 & 0.00000 & 30.2651 & CJ2\\
    \hline
    \multicolumn{1}{c}{ 08/23/2013   (BO043L) } \\
    \hline
    1.54091 & 0.817234 & -98.3806 & 5.08116 & 0.201800 & 31.5805 & core\\
    0.170572 & 50.6366 & 49.0882 & 19.1943 & 0.00000 & 42.1254 & C1\\
    0.326132 & 29.1062 & 50.3968 & 27.1114 & 0.00000 & 44.0108 & C2\\
    0.124930 & 22.4459 & 48.1116 & 12.1035 & 0.213216 & 15.4242 & J7\\
    0.652417 & 11.5222 & 49.9473 & 7.21172 & 0.305156 & 32.5915 & J10\\
    0.904259 & 2.96773 & 56.1347 & 6.74834 & 0.232570 & 30.3491 & C3\\
    0.163622 & 16.3465 & -125.763 & 7.08633 & 0.00000 & 71.0185 & CJ2 \\
    \hline \hline
    \end{tabular}
\end{table*}

\section{Montage of images}
\label{sec:appendix:montage}

\begin{figure*}[h!]
\begin{center}
\includegraphics[width=0.8\linewidth]{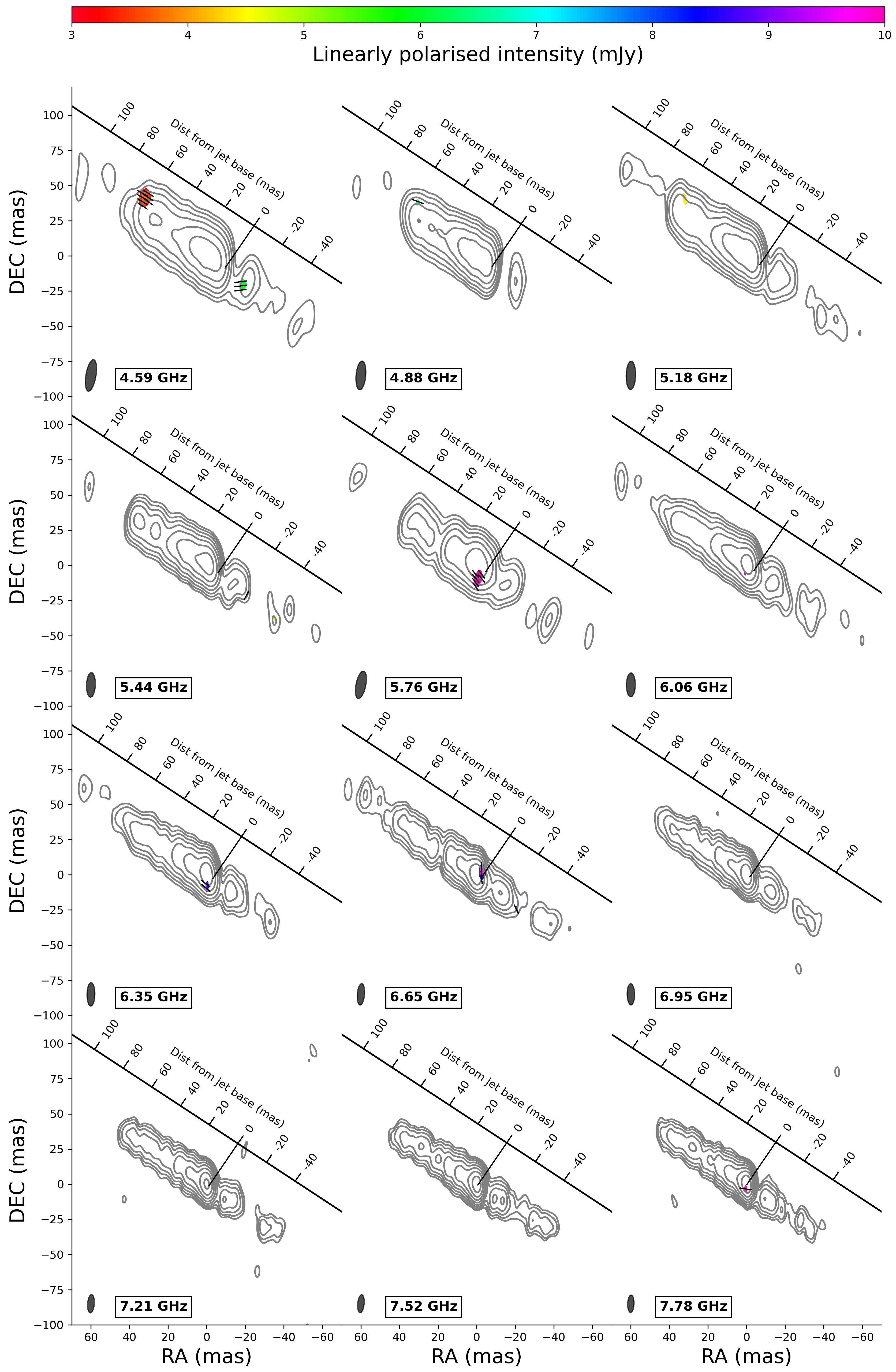}
\caption{Montage of all images from epoch 1. The contour levels are at $noise \times 2^{n}$, where $n=1, 2, 3, 4, ...$. For frequency bands with an asterisk next to the frequency, no reliable EVPA calibration could be performed. The vectors show the EVPA-corrected electric field directions but are not corrected for Faraday rotation. }
\label{fig:montage1}
\end{center}
\end{figure*}

\begin{figure*}[h!]
\begin{center}
\includegraphics[width=0.8\linewidth]{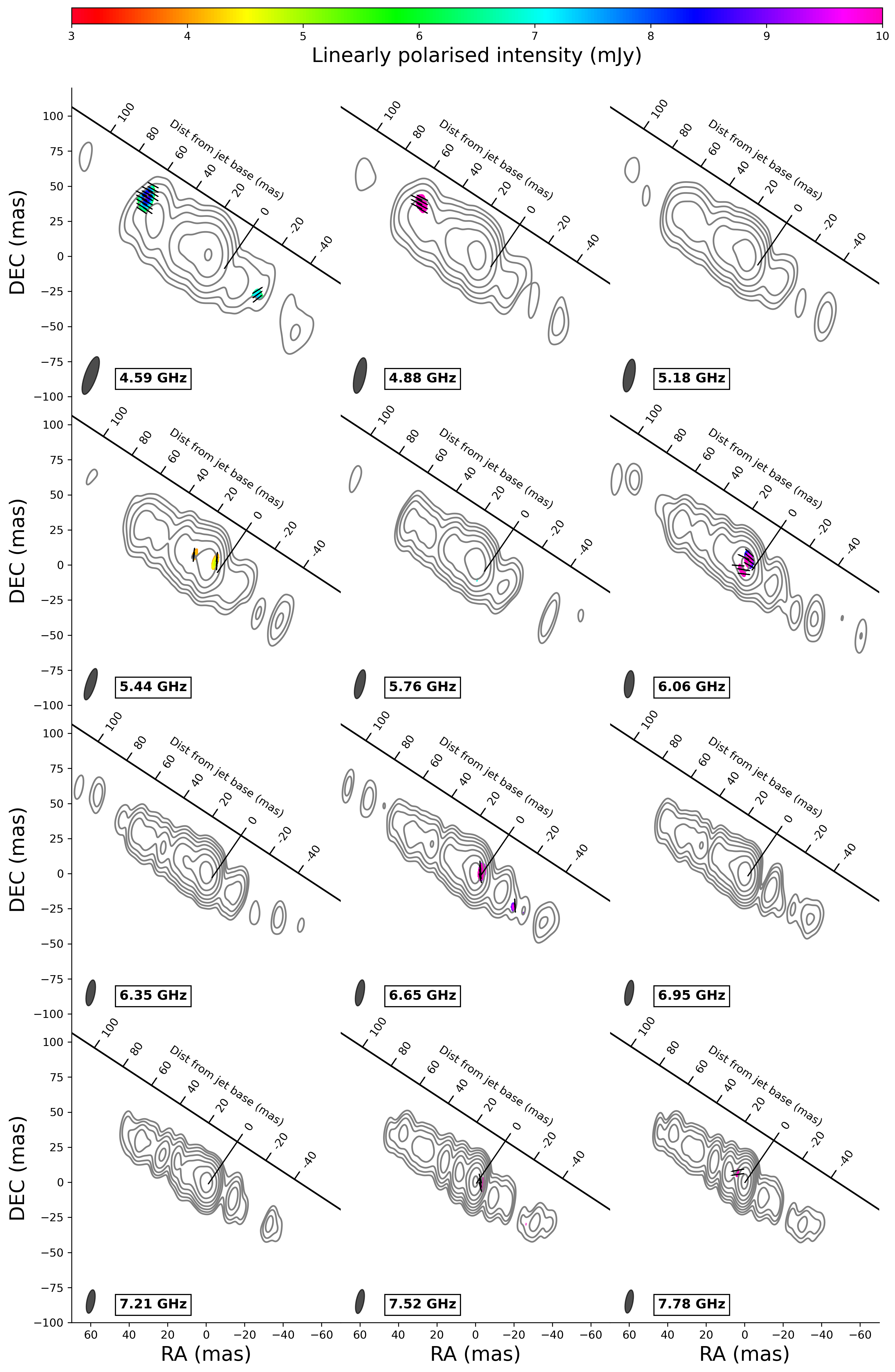}
\caption{Same as Figure \ref{fig:montage1} but for epoch 2.}
\label{fig:montage2}
\end{center}
\end{figure*}

\begin{figure*}[h!]
\begin{center}
\includegraphics[width=0.8\linewidth]{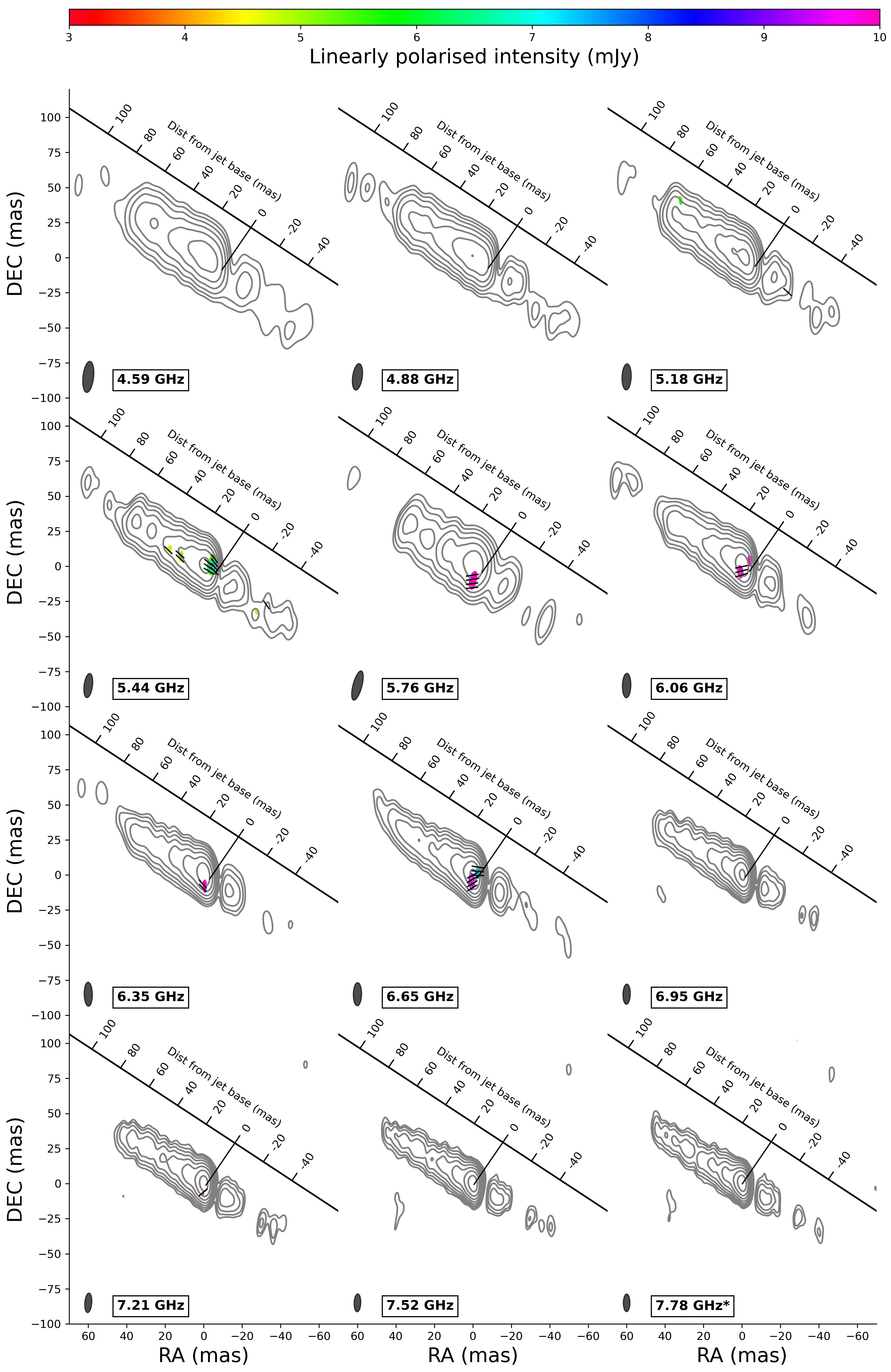}
\caption{Same as Figure \ref{fig:montage1} but for epoch 3.}
\label{fig:montage3}
\end{center}
\end{figure*}

\begin{figure*}[h!]
\begin{center}
\includegraphics[width=0.8\linewidth]{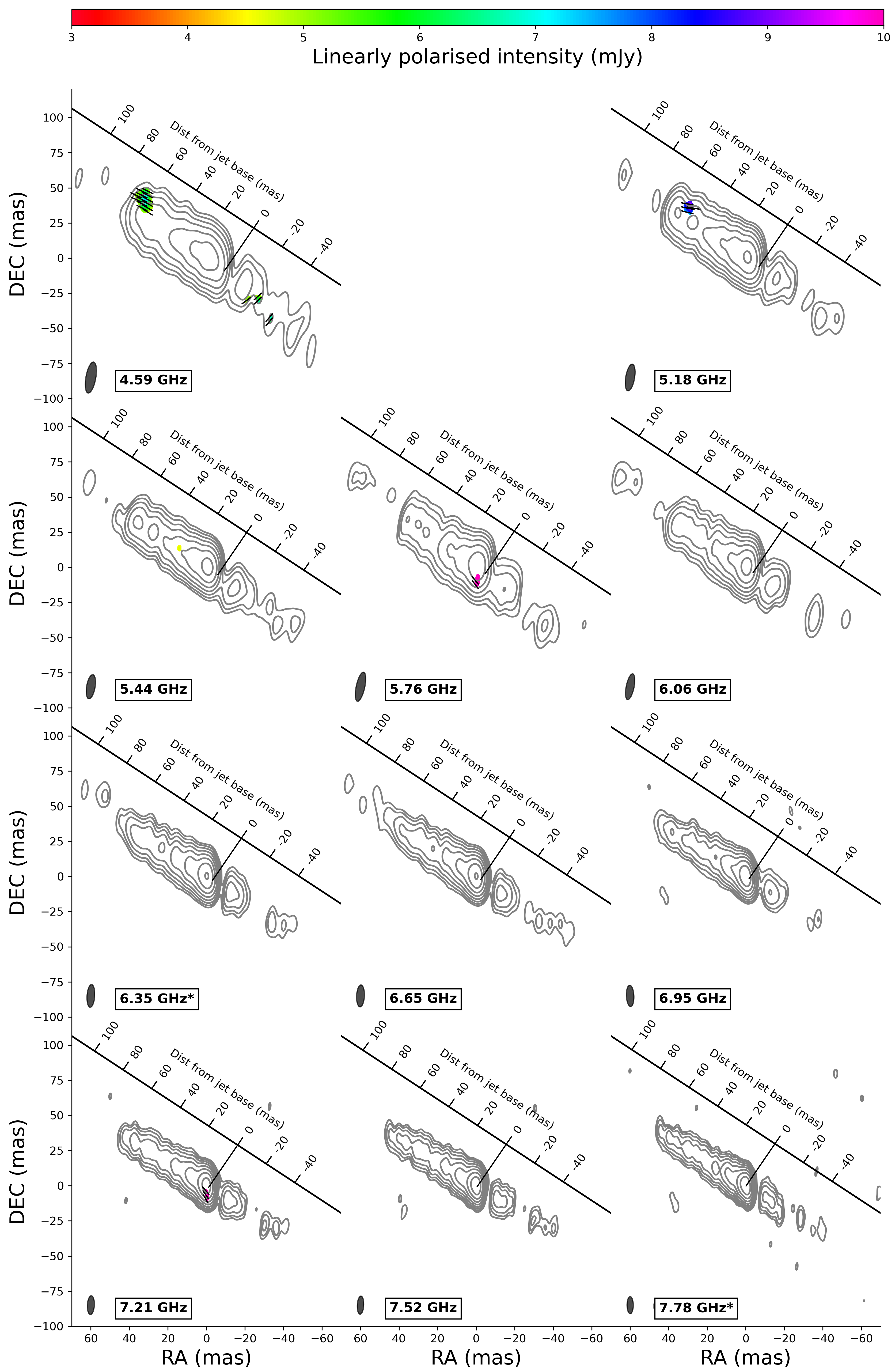}
\caption{Same as Figure \ref{fig:montage1} but for epoch 4.}
\label{fig:montage4}
\end{center}
\end{figure*}

\begin{figure*}[h!]
\begin{center}
\includegraphics[width=0.8\linewidth]{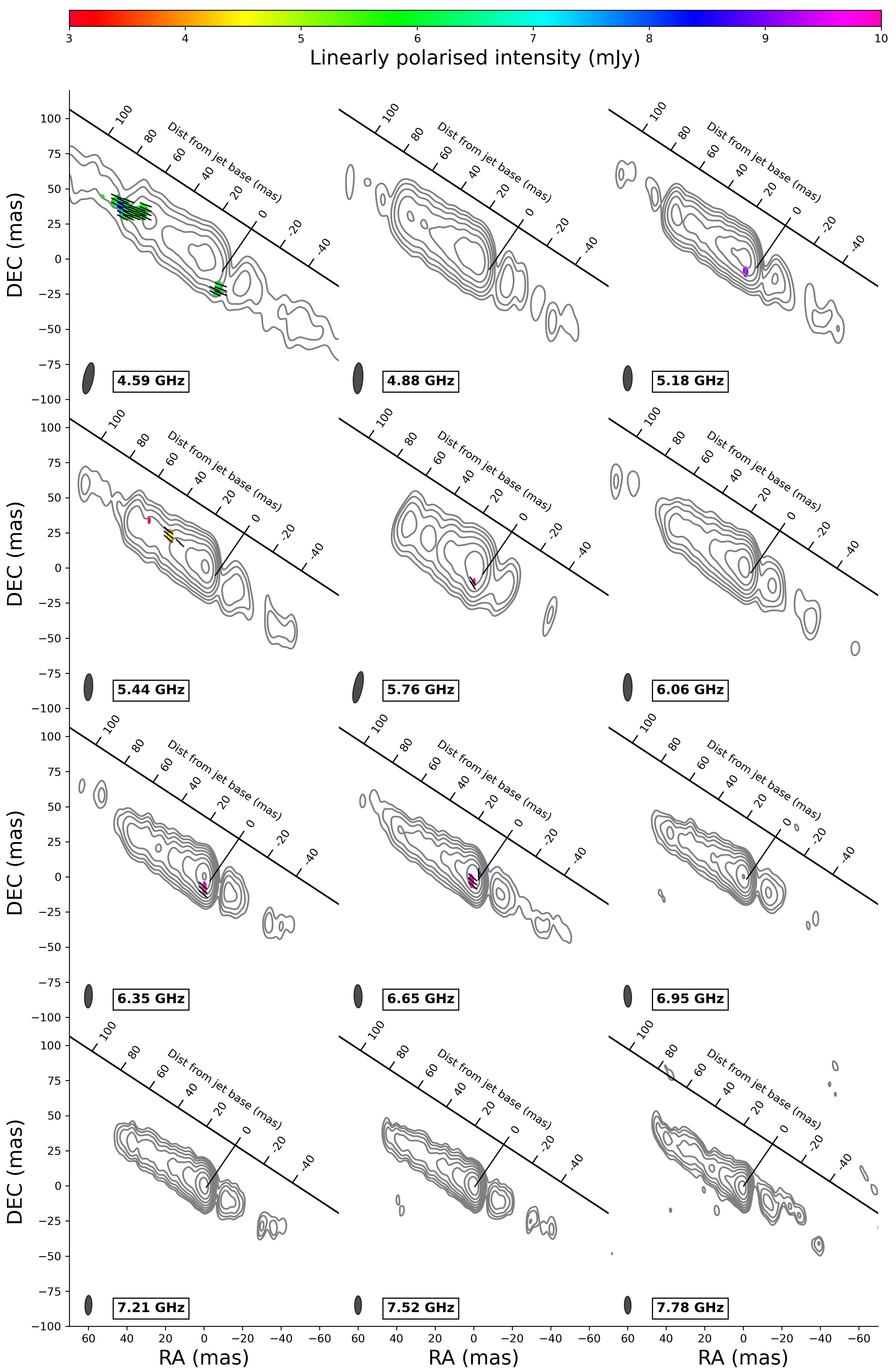}
\caption{Same as Figure \ref{fig:montage1} but for epoch 5.}
\label{fig:montage5}
\end{center}
\end{figure*}

\begin{figure*}[h!]
\begin{center}
\includegraphics[width=0.8\linewidth]{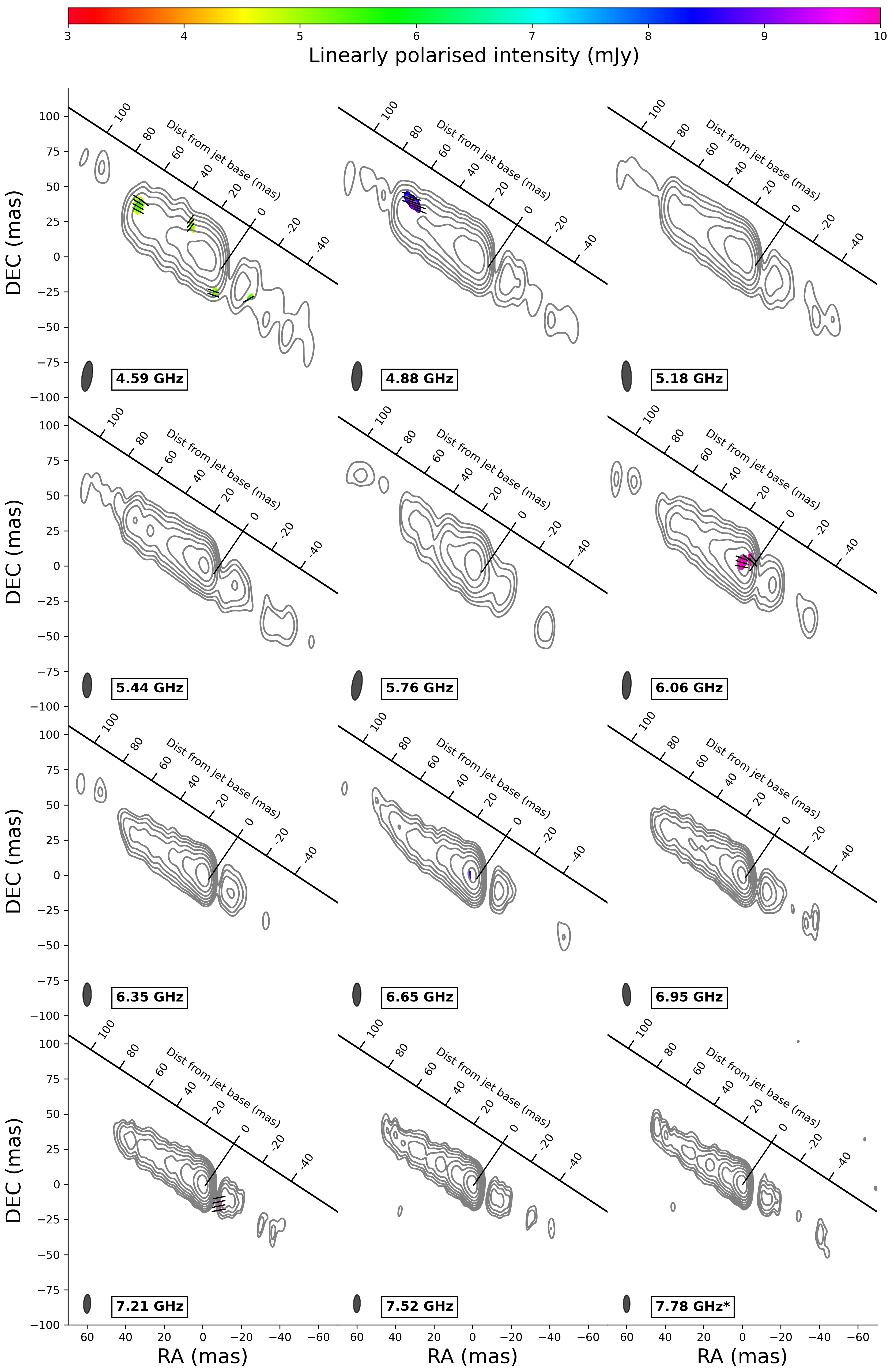}
\caption{Same as Figure \ref{fig:montage1} but for epoch 6.}
\label{fig:montage6}
\end{center}
\end{figure*}

\begin{figure*}[h!]
\begin{center}
\includegraphics[width=0.8\linewidth]{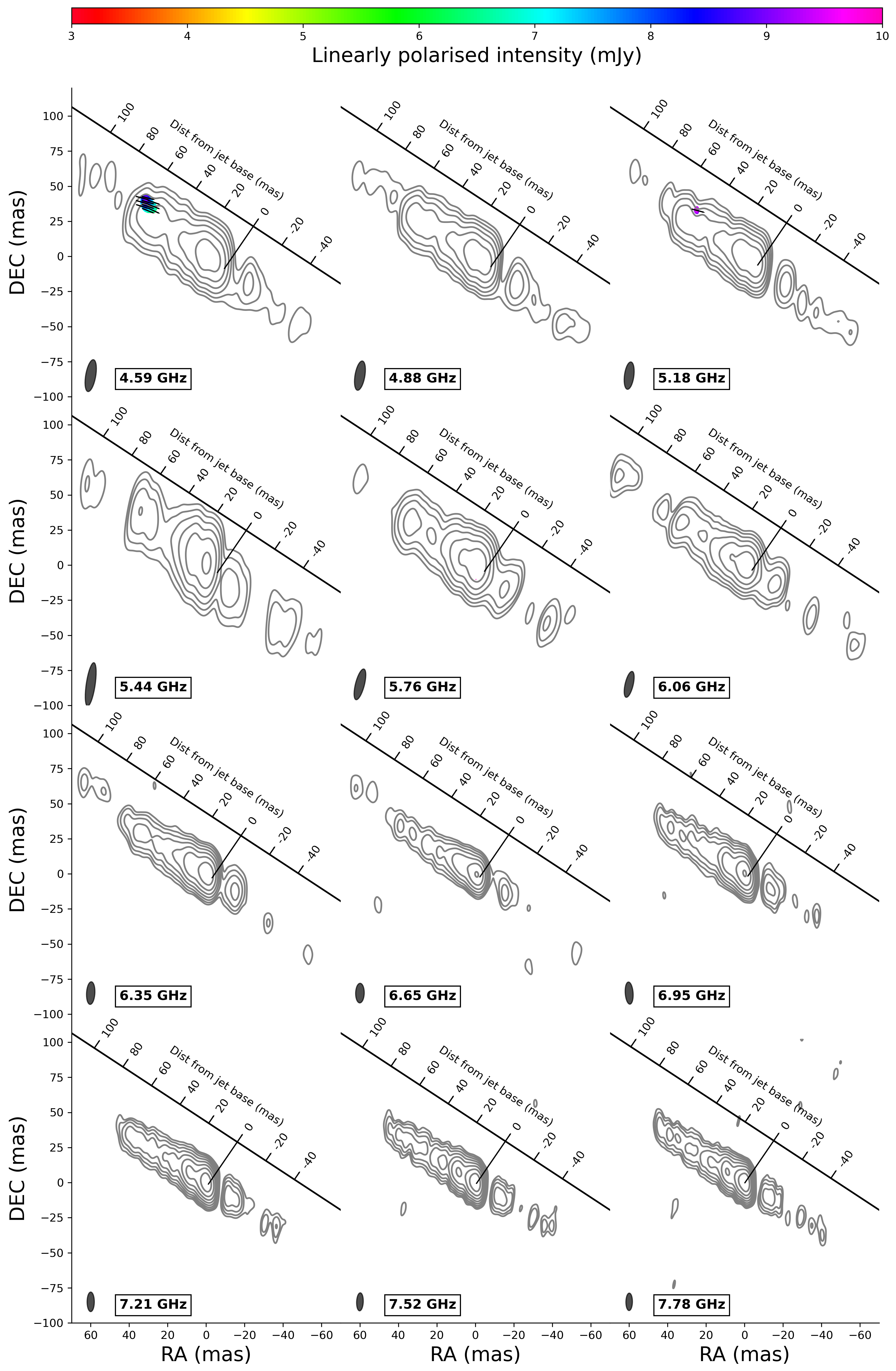}
\caption{Same as Figure \ref{fig:montage1} but for epoch 12.}
\label{fig:montage12}
\end{center}
\end{figure*}

\newpage
\section{Nature of CJ2 Component: Extended Discussion}
\label{sec:appendix:natureOfCJ2}
\begin{figure*}[h]
\begin{center}
\includegraphics[width=0.8\linewidth]{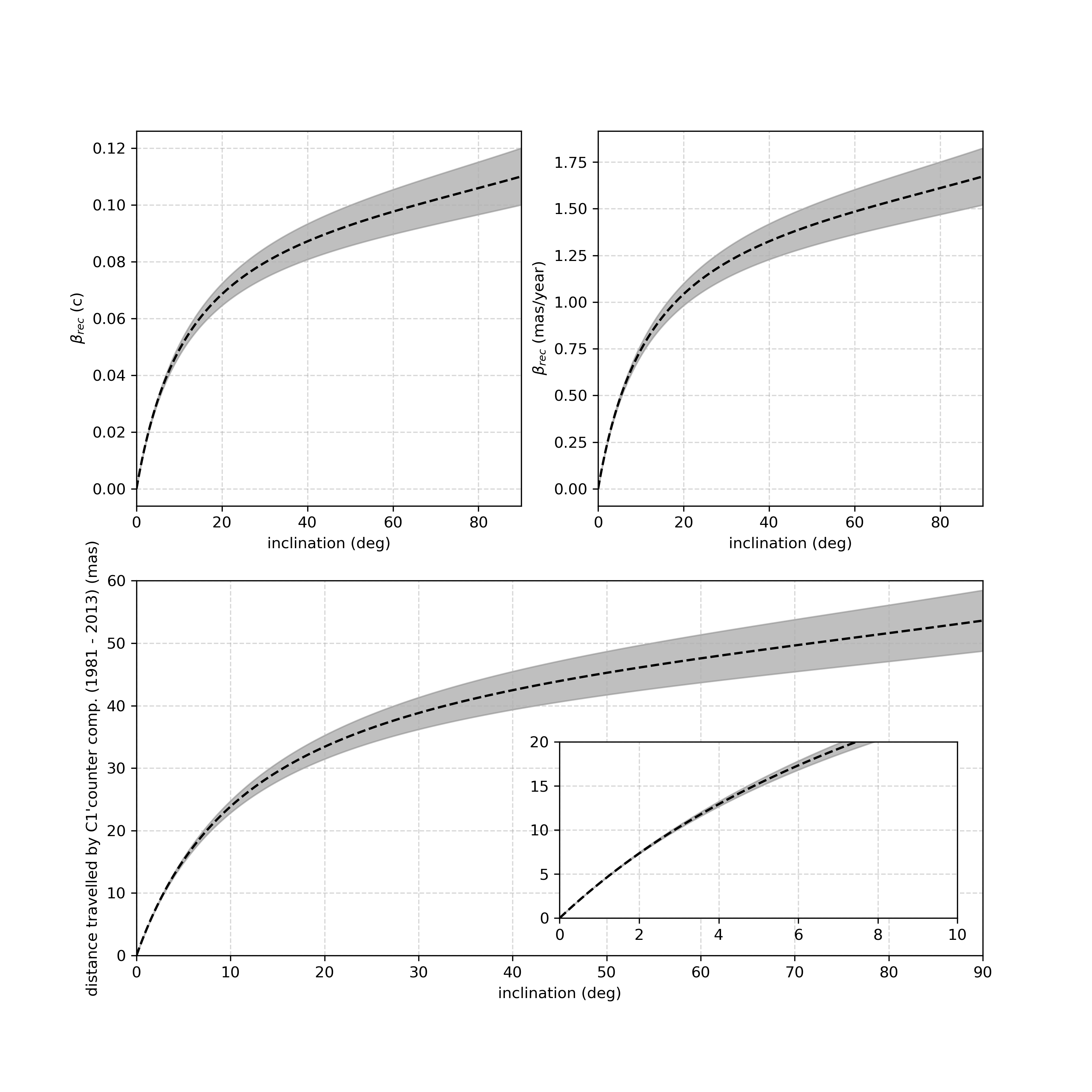}
\caption{In the top two panels, we show the predicted apparent speed of 
C1's equivalent receding jet ejection for various inclination angles. In the bottom plot, we show the expected distance traveled downstream by C1's receding jet equivalent, since C1's estimated ejection time.}
\label{fig:C1VsCJ2}
\end{center}
\end{figure*}

\begin{figure*}[h]
\begin{center}
\includegraphics[width=0.8\linewidth]{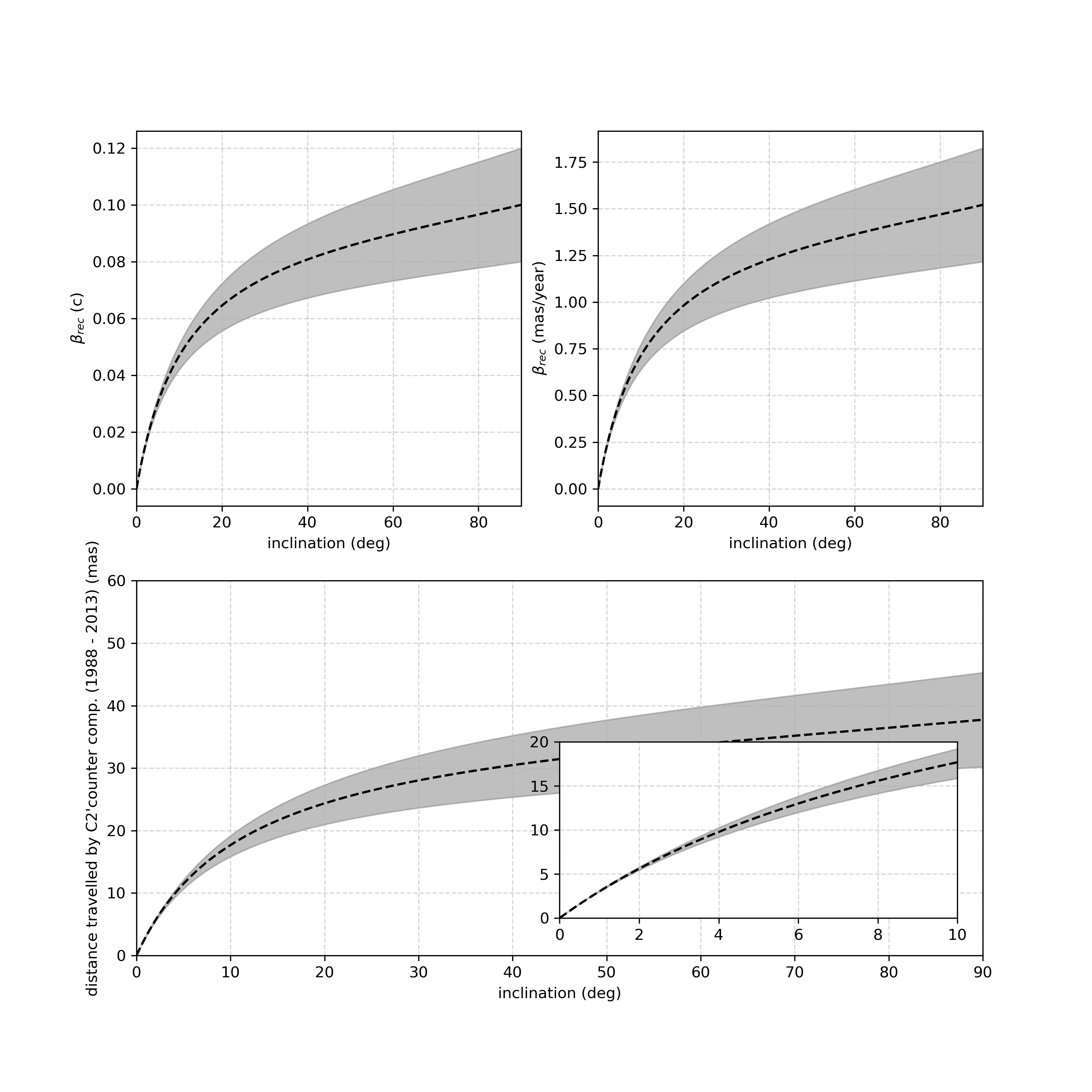}
\caption{Same as Figure \ref{fig:C1VsCJ2}, but for the C2 component.}
\label{fig:C2VsCJ2}
\end{center}
\end{figure*}

In this section, using equations that define the proper motion and brightness ratio of simultaneously launched jet components, we rule out the possibility of CJ2 being the associated receding jet ejection of either C1 or C2 components. For an approaching component with an apparent speed of $\beta_{app}$, the apparent speed of its corresponding receding component along the receding jet is given by \\ 

\begin{equation}
    \beta_{rec} = \frac{\beta \sin(\theta)}{1 + \beta \sin(\theta)},
    \label{eq:appendix1}
\end{equation}
where $\theta$ is the inclination and $\beta$ is the  intrinsic speed. The speed can also be written in terms of the approaching jet speed using Equation \ref{Eq:inclinationConstrain1}. Using Equations \ref{eq:appendix1} and \ref{Eq:inclinationConstrain1}, along with the known $\beta_{app}$ and $T_{ejection}$ of C1 and C2 (from Table \ref{tab:kinematics}), we plot the corresponding $\beta_{rec}$ and the total distance traveled since ejection, in Figure \ref{fig:C1VsCJ2} and Figure \ref{fig:C2VsCJ2}. 
The CJ2 component has an apparent speed of $0.01 \pm 0.01$\,c with $<5$\, mas distance traveled during 20 years. From the above Figures, for CJ2 to be associated with the C1 or C2 component would require an inclination  $<2^{\circ}$. \\

\section{Stacked montage}
\label{sec:appendix:montageStacked}

\begin{figure*}[h!]
\begin{center}
\includegraphics[width=0.8\linewidth]{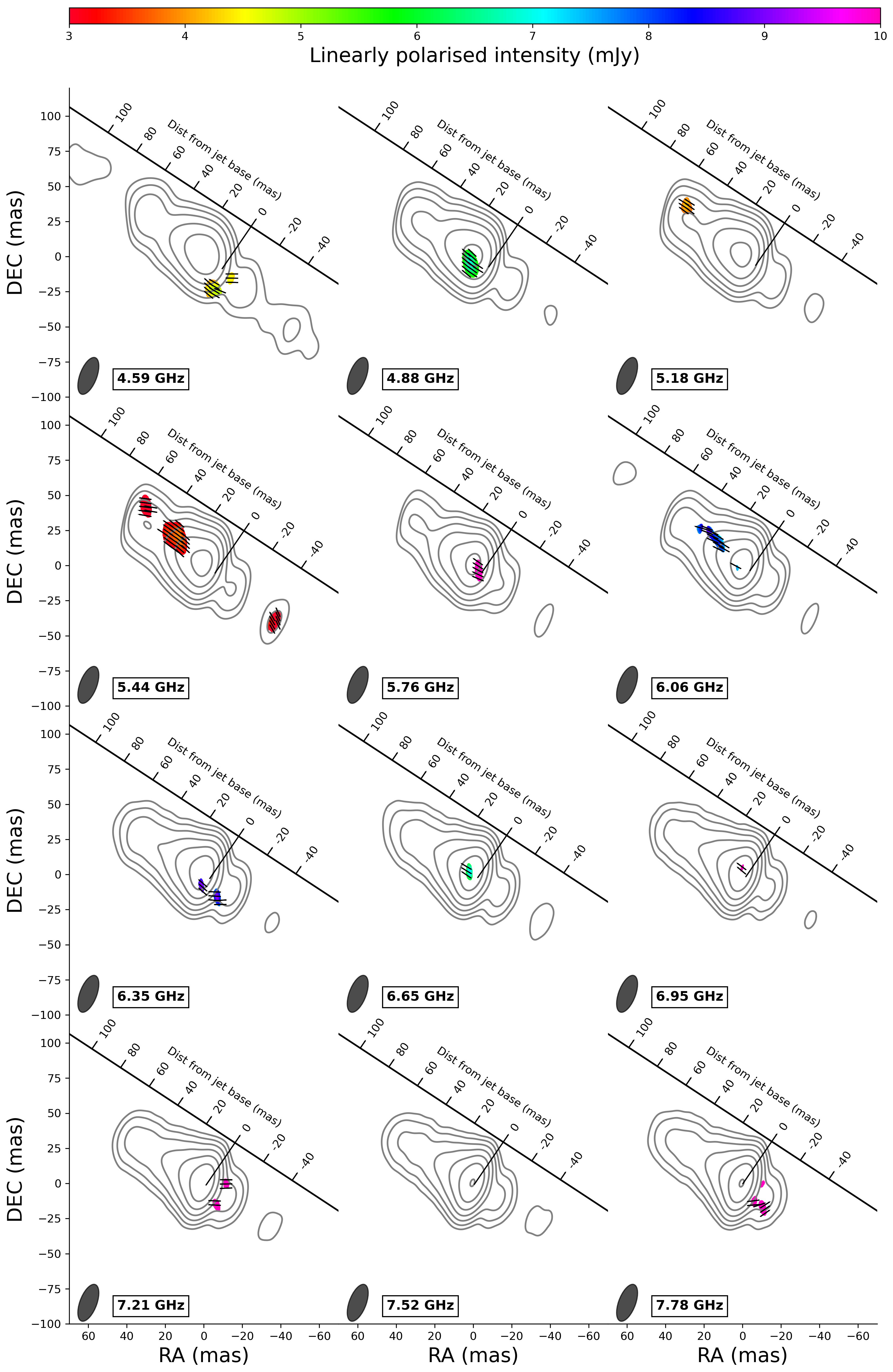}
\caption{Montage of all frequencies having stacked across the different epochs. The contour levels are at $noise \times \sqrt{2}^{n}$, where $n=2, 4, 6, 8, ...$. The vectors show the EVPA-corrected electric field directions but are not corrected for Faraday rotation.  }
\label{fig:montageStacked}
\end{center}
\end{figure*}

\section{Opening angle}
\label{sec:appendix:openingAngle}
In this section, we derive the jet opening angle by assuming the star to be crossing the jet stream in Centaurus A at a distance of 22mas from the central black hole of mass M. Assuming the star to be in a circular orbit, we get its angular velocity using the below equation.

\begin{equation}
    \omega = \sqrt{\frac{GM}{r^{3}}},
\end{equation}
where G is the gravitational constant, and r is the 22mas measured on the sky plane de-projected for inclination effects. Using the above equation and the inclination angle constraint ($\le 25^{\circ}$), we find the jet opening angle to be $\le 0.6$ degrees for a 20 year jet crossing time.


\bibliography{sample631}{}
\bibliographystyle{aasjournal}



\end{document}